\documentclass[sigconf]{acmart}
\usepackage{threeparttable} 
\usepackage{microtype}
\usepackage{tipa}
\usepackage{graphicx}
\usepackage{subfigure}
\usepackage{balance}
\usepackage{booktabs} 
\usepackage{amsthm}
\usepackage{array}
\usepackage{graphicx}
\usepackage{clrscode}
\usepackage{subfigure}
\usepackage{multirow}
\usepackage{multicol}
\usepackage{float}
\usepackage{color}
\usepackage{xcolor}
\usepackage{amsopn}
\usepackage{mathrsfs}
\usepackage{mathtools}
\usepackage{amsmath}
\usepackage{booktabs}
\usepackage{arydshln}
\usepackage{hyperref}
\usepackage{blkarray}
\usepackage{enumerate}
\usepackage{courier}
\usepackage{mathrsfs}
\usepackage{rotating}
\usepackage{bm}
\usepackage{subfigure}
\usepackage{array}
\usepackage{ragged2e}
\usepackage{hyperref}
\usepackage{amsmath}
\usepackage{listings}
\usepackage[customcolors]{hf-tikz}
\usepackage{xspace}
\usepackage[misc]{ifsym}
\newcommand{\ignore}[1]{}

\usepackage{adjustbox} 
\usepackage{enumitem}
\newcommand{\tabincell}[2]{\begin{tabular}{@{}#1@{}}#2\end{tabular}} 
\usepackage{comment}

\usepackage[ruled,linesnumbered]{algorithm2e}

\SetKwComment{comment}{ $triangleright$ \ }{}

\newcommand{\nummodel}{73\xspace}
\newcommand{\numdataset}{28\xspace}
\newcommand{\modelname}{RecBole\xspace}
\newcommand{\weblink}{\url{https://recbole.io/}}

\AtBeginDocument{%
  \providecommand\BibTeX{{%
    \normalfont B\kern-0.5em{\scshape i\kern-0.25em b}\kern-0.8em\TeX}}}

\setcopyright{acmcopyright}
\copyrightyear{2021}
\acmYear{2021}
\setcopyright{acmlicensed}\acmConference[CIKM '21]{Proceedings of the 30th ACM International Conference on Information and Knowledge Management}{November 1--5, 2021}{Virtual Event, QLD, Australia}
\acmBooktitle{Proceedings of the 30th ACM International Conference on Information and Knowledge Management (CIKM '21), November 1--5, 2021, Virtual Event, QLD, Australia}
\acmPrice{15.00}
\acmDOI{10.1145/3459637.3482016}
\acmISBN{978-1-4503-8446-9/21/11}

\begin{document}
\fancyhf{} 
\fancyhead[C]{Submission to ACM CIKM Resource Track} 
\fancyfoot[C]{\thepage}

\title{\modelname: Towards a Unified, Comprehensive and Efficient\\ Framework for Recommendation Algorithms}
\author{Wayne Xin Zhao$^{1,2}$, Shanlei Mu$^{1,3,\#}$, Yupeng Hou$^{1,2,\#}$, Zihan Lin$^{1,3}$, Yushuo Chen$^{1,2}$, Xingyu Pan$^{1,3}$, Kaiyuan Li$^{4}$, Yujie Lu$^{7}$, Hui Wang$^{1,3}$, Changxin Tian$^{1,3}$, Yingqian Min$^{1,3}$, Zhichao Feng$^{4}$, Xinyan Fan$^{1,2}$, Xu Chen$^{1,2,*}$, Pengfei Wang$^{4,*}$, Wendi Ji$^{5}$, Yaliang Li$^{6}$, Xiaoling Wang$^{5}$, Ji-Rong Wen$^{1,2,3}$}
\thanks{$^*$ Xu Chen (successcx@gmail.com) and Pengfei Wang (wangpengfei@bupt.edu.cn) are corresponding authors.}
\thanks{$\#$ Both authors contributed equally to this work.}
\affiliation{\institution{$^1$Beijing Key Laboratory of Big Data Management and Analysis Methods} \country{}} 
\affiliation{\institution{\{$^2$Gaoling School of Artificial Intelligence, $^3$School of Information\} Renmin University of China} \country{}}
\affiliation{\institution{$^4$Beijing University of Posts and Telecommunications, $^5$East China Normal University, $^6$Alibaba, $^7$Liaoning University} \country{}}

\begin{abstract}
In recent years, there are a large number of recommendation algorithms proposed in the literature, from traditional collaborative filtering to deep learning algorithms. 
However, the concerns about how to standardize  open source implementation of recommendation algorithms continually increase in the research community. 
In the light of this challenge, we propose a unified, comprehensive and efficient recommender system library called \emph{\modelname} (pronounced as [r\textepsilon k'bo\textipa{U}l\textschwa r]), which provides a unified framework to develop and reproduce recommendation algorithms for research purpose.
In this library, we implement \nummodel recommendation models on \numdataset  benchmark datasets, covering the categories of general recommendation, sequential recommendation, context-aware recommendation and knowledge-based recommendation. We implement the \modelname library based on PyTorch, which is one of  the most popular deep learning frameworks.
Our library is featured in many aspects, including general and extensible data structures, 
comprehensive benchmark models and datasets, efficient GPU-accelerated execution, and extensive and standard evaluation protocols. 
We provide a series of auxiliary functions, tools, and scripts to facilitate  the use of this library, such as automatic parameter tuning and break-point resume.
Such a framework is useful to standardize the implementation and evaluation of recommender systems. 
The project and documents are released at \weblink.
\end{abstract}

\maketitle
\fancyhead{}

\section{Introduction}
In the era of big data, recommender systems are playing a key  role 
in tackling information overload, which largely improve the user experiences in a variety of applications, ranging from e-commerce, video sharing to healthcare assistant and on-line education.
The huge business value makes recommender systems become a longstanding research topic, with a large number of new models proposed each year~\citep{DLRS-survey}.
With the rapid growth of recommendation algorithms, these algorithms are usually developed under  different platforms or  frameworks. Typically, an experienced  researcher often  finds it difficult to implement the compared baselines in a unified way or framework. Indeed, many common components or procedures of these recommendation algorithms  are  duplicate or highly similar, which should be reused or extended. 
Besides, we are aware that 
there is an increasing concern about model reproducibility in the research community. 
Due to some reasons, many published recommendation algorithms still lack public  implementations. Even with open source code, many details are implemented inconsistently (e.g., with different loss functions or optimization strategies) by different developers.  
There is a need to re-consider the implementation of recommendation algorithms in a unified way.

In order to alleviate the above issues, we initiate a project  to provide a unified framework for developing recommendation algorithms.
We implement an open source recommender system library, called \emph{\modelname} (pronounced as [r\textepsilon k'bo\textipa{U}l\textschwa r]) \footnote{Bole was a  famous Chinese judge of horses in Spring and Autumn period, who was the legendary inventor of equine physiognomy (``\emph{judging a horse's qualities from appearance}''). Bole is frequently associated with the fabled \emph{qianlima} (a Chinese word) ``thousand-\emph{li} horse'', which was supposedly able to gallop one thousand \emph{li} (approximately 400 km) in a single day. Read more details about Bole at the wikipedia page via the link: \url{https://en.wikipedia.org/wiki/Bo_Le}. Here, we make an analogy between identifying qianlima horses and making good recommendations.}.
Based on this library, we would like to enhance the  reproducibility of existing models and ease the developing process of new  algorithms.  Our work is also useful to standardize the evaluation protocol of recommendation algorithms. 
Indeed, a considerable number of recommender system libraries have been released in the past decade~\citep{Librec,MyMediaLite,NeuRec,ReChorus,daisyRec}. These works have largely advanced the progress of open source recommender systems. 
Many libraries have made continuous improvement with increasingly added features.
We have extensively surveyed these libraries and broadly fused their merits into \modelname. 
To summarize,  the key features and capabilities 
 of our \modelname library are summarized in the following five aspects:

$\bullet$ Unified recommendation framework. We adopt PyTorch~\citep{PyTorch} to develop the entire recommender system library, since it is one of the most popular deep learning frameworks, especially in the research community.
As three core components of our library, we design and develop data modules, model modules, and evaluation modules, and encapsulate many common components, functions or procedures shared by different recommendation algorithms. 
In our library, for reusing existing models, one can easily compare different recommendation algorithms with built-in evaluation protocols via simple  yet  flexible configuration; for developing new models, one  only needs to focus on a small number of interface functions, so that common parts  can be reused and implementation details are made transparent to the users. 

$\bullet$ General and extensible data structure. For unified algorithm development, we implement the supporting data structures at two levels. At the user level, we introduce  \emph{atomic files} to format the input of mainstream recommendation tasks in a flexible way.
The proposed atomic files are able to characterize the input of four kinds of mainstream recommendation tasks. 
At the algorithm level, we introduce a general data structure \textsf{Interaction} to unify the internal data representations tailored to GPU-based environment.  The design of \textsf{Interaction} is particularly convenient to develop new algorithms with supporting mechanisms or functions, e.g.,  fetching the data  by referencing feature name. We implement  \textsf{Dataset} and  \textsf{DataLoader} (two python classes) to automate the entire data flow, which greatly reduces the efforts for developing new models.

\begin{figure}[t]
\centering
\setlength{\fboxrule}{0.pt}
\setlength{\fboxsep}{0.pt}
\fbox{
\includegraphics[width=1.\linewidth]{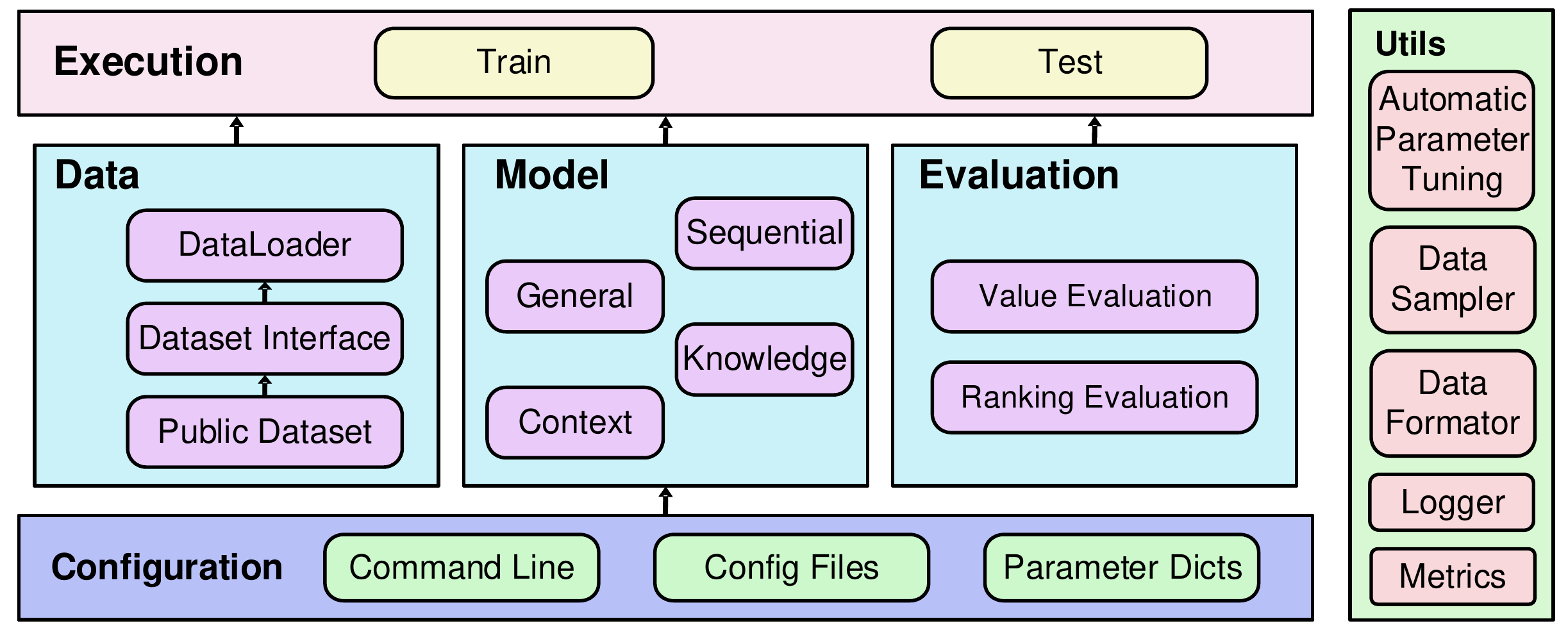}
}
\caption{The overall framework of our library \modelname.}
\vspace{-0.1cm}
\label{fw}
\end{figure}

$\bullet$ Comprehensive benchmark models and datasets.
So far, we have implemented \nummodel recommendation algorithms, covering 
the categories of general recommendation, sequential recommendation, context-aware recommendation and knowledge-based recommendation.
Besides traditional recommendation algorithms, we  incorporate a large number of neural algorithms proposed in recent years. 
We provide flexible supporting mechanisms via the configuration files or command lines to run, compare and test these algorithms. 
We also implement rich auxiliary  functions to use these models, including automatic parameter tuning and break-point resume. 
To construct a reusable benchmark, we incorporate \numdataset commonly used datasets for evaluating recommender systems.
With original dataset copies, a user can simply transform the data into a form that can be used in our library with the provided preprocessing tools or scripts.
More datasets and methods will be incorporated into our library.

$\bullet$ Efficient GPU-accelerated  execution. We design and implement a number of efficiency optimization techniques that are tailored to the GPU environment. As two major sources of time costs, both the model training and  testing are accelerated with GPU-oriented implementations. 
For model test,  a special acceleration strategy is proposed to improve the efficiency of the full ranking for top-$K$ item recommendation. We   convert the  top-$K$ evaluation for all the users into the computation based on a unified matrix form. With this matrix form, we can utilize the GPU-version $\textsf{topk()}$  function in PyTorch to directly optimize  the top-$K$ finding procedure. Furthermore, such a matrix form is particularly convenient for generating the recommendations and computing the evaluation metrics. We empirically show that it  significantly reduces the time cost of the straightforward implementation without our acceleration strategy.  

$\bullet$ Extensive and standard evaluation protocols.  Our library supports a series of widely adopted evaluation protocols for testing and comparing  recommendation algorithms. 
It incorporates the various  evaluation settings discussed in \citep{comparison}.
Specially, we implement different combinations of item sorting (i.e., how to sort the items before data splitting) and data splitting (i.e., how to derive the train/validation/test sets) for deriving the evaluation sets. We also consider both full ranking and sample-based ranking, which is recently a controversial issue in the field of recommender system~\citep{sampledmetrics}.  We encapsulate four basic interfaces (namely \emph{Group}, \emph{Split}, \emph{Order} and \emph{NegSample})  to support the above evaluation protocols, which is flexible to include other evaluation settings. We provide a few commonly used evaluation settings (e.g., ratio-based splitting plus random ordering for dataset splitting), which integrates the alternative settings  of the above four factors. Our library provides a possibility to evaluate models under different evaluation settings.

\renewcommand\arraystretch{1.2}
\begin{table}[t]
\centering
\caption{Collected datasets in our library \modelname.}
\vspace{-0.cm}
\label{tab:public-dataset}
\begin{threeparttable}  
\begin{tabular}
{p{2.5cm}<{\centering}||p{1.4cm}<{\centering}|p{1.4cm}<{\centering}|p{1.7cm}<{\centering}}
\hline
Dataset & \#Users & \#Items & \#Interactions\\\hline\hline
MovieLens  & - & - & - \\
Anime & 73,515 & 11,200 & 7,813,737 \\
Epinions & 116,260 & 41,269 & 188,478\\
Yelp & 1,968,703 & 209,393 & 8,021,122  \\
Netflix & 480,189 & 17,770 & 100,480,507 \\
Book-Crossing & 105,284 & 340,557 & 1,149,780 \\
Jester & 73,421 & 101 & 4,136,360  \\
Douban & 738,701 & 28 & 2,125,056\\
Yahoo Music & 1,948,882 & 98,211 & 11,557,943 \\
KDD2010 & - & - & -  \\
Amazon & - & - & -  \\
Pinterest & 55,187 & 9,911 & 1,445,622 \\
Gowalla & 107,092 & 1,280,969 & 6,442,892 \\
Last.FM & 1,892 & 17,632 & 92,834  \\
DIGINETICA & 204,789 & 184,047 & 993,483  \\
Steam & 2,567,538 & 32,135 & 7,793,069 \\
Ta-Feng & 32,266 & 23,812 & 817,741 \\
FourSquare & - & - & - \\
Tmall & 963,923 & 2,353,207 & 44,528,127 \\
YOOCHOOSE & 9,249,729 & 52,739 & 34,154,697 \\
Retailrocket & 1,407,580 & 247,085 & 2,756,101 \\
LFM-1b & 120,322 & 3,123,496 & 1,088,161,692 \\
Criteo & - & - & 45,850,617 \\
Avazu & - & - & 40,428,967 \\
iPinYou & 19,731,660 & 163 & 24,637,657 \\
Phishing websites & - & - & 11,055 \\
Adult & - & - & 32,561 \\
MIND & - & - & -\\
\hline
\end{tabular}
\begin{tablenotes}    
   \footnotesize              
   \item[1] ``-'' means the dataset is either composed of many small subsets (e.g., Amazon, KDD2010), so that we refer the readers to our website for more detail statistics or the dataset is based on the features (e.g., Criteo, Avazu), instead of  users and items. 
\end{tablenotes}            
\end{threeparttable}
\vspace{-0.2cm}  
\end{table}

\section{The Library --- \modelname}
The overall framework of our library \modelname is presented in Figure~\ref{fw}.
The bottom part is the configuration module, which helps users to set up the experimental environment (e.g., hyperparameters and running details).
The data, model and evaluation modules are built upon the configuration module, which forms the core code of our library.
The execution module is responsible for running and evaluating the model based on specific settings of the environment.
All the auxiliary functions are collected in the utility module, including automatic parameter tuning, logger and evaluation metrics.
In the following, we briefly present the designs of three core modules, and more details can be found in the library documents.

\subsection{Data Module}
A major development guideline of our library is to make the code highly 
self-contained and unified. For this purpose, data module is indeed the most important part that supports the entire library by  providing fundamental data structures and functions.

\subsubsection{The Overall Data Flow}
For extensibility and reusability, our data module designs an elegant data flow that transforms raw data into the model input.

The overall data flow can be described as follows: \underline{raw input} $\rightarrow$ \underline{atomic files}  $\rightarrow$ \underline{\textsf{Dataset}$_{DataFrame}$}  $\rightarrow$ \underline{\textsf{Dataloader}$_{Interaction}$} $\rightarrow$ \underline{algorithms}.
The implementation of  class \textsf{Dataset} is mainly based on the primary data structure of  \textsf{pandas.DataFrame} in the library of pandas, and the implementation of  class   \textsf{Dataloader}  is based on a general internal data structure called  \textsf{Interaction}.

Our data flow involves two special data forms, which are oriented to users and algorithms, respectively.  For data preparation, we introduce and define six \emph{atomic file types} (having the same or similar file format) for unifying the input at the user level. While, for internal data representations, we introduce and implement a flexible data  structure  \textsf{Interaction} at the algorithm level. 
The atomic files are able to characterize most forms of the input data  required by different recommendation tasks, and the \textsf{Interaction} data structure provides a unified internal data representation for different recommendation algorithms.

In order to help users transform raw input into atomic files,  we have collected more than \numdataset commonly used datasets and released the corresponding conversion tools, which makes it quite convenient to start with our library. We present the statistics of these datasets in Table~\ref{tab:public-dataset}.
During the transformation step from atomic files to class \textsf{Dataset}, we provide many useful functions that support a series of preprocessing steps in recommender systems, such as $k$-core data filtering and  missing value imputation. We present the  functions supported by class \textsf{Dataset} in Table~\ref{tb-dataset}.  

\renewcommand\arraystretch{1.2}
\begin{table}[t]
\centering
\caption{The functions supported by class \textsf{Dataset}.}
\vspace{-0.cm}
\label{tb-dataset}
\begin{tabular}{p{3.cm}<{\centering}||p{4.5cm}<{\centering}}
\hline
Function & Description\\
\hline
\hline
\textsf{\_filter\_by\_inter\_num} & frequency based user/item filtering\\ 
\textsf{\_filter\_by\_field\_value} & value based filtering\\ 
\textsf{\_remap\_ID} & map the features to IDs \\
\textsf{\_fill\_nan} & missing value imputation \\
\textsf{\_set\_label\_by\_threshold} & generate interaction labels\\
\textsf{\_normalize} & normalize the features\\
\textsf{\_preload\_weight\_matrix} & initialize embedding tables \\
\hline
\end{tabular}
\end{table}

\subsubsection{Atomic Files}

So far, our library introduces six atomic file types, which are served as basic components for characterizing the input of various recommendation tasks.  
In the literature, there is a considerable number of recommendation tasks. 
We try to summarize and unify the most basic input forms for mainstream  recommendation tasks. 
Note that these files are only functionally different while their formats are rather similar. 
The details of these atomic files are summarized in Table~\ref{tb-atom}.

\renewcommand\arraystretch{1.2}
\begin{table}[t]
\centering
\caption{Summarization of the atomic files.}
\vspace{-0.cm}
\begin{threeparttable} 
\begin{tabular}{p{1.cm}<{\centering}||p{1.5cm}<{\centering}|p{4.2cm}<{\centering}}
\hline
Suffix&Data types&Content\\ \hline\hline
\textsc{.inter}&all types&User-item interaction\\ 
\textsc{.user}&all types&User feature \\
\textsc{.item}&all types&Item feature\\ 
\textsc{.kg}&int&Triplets in a knowledge graph \\ 
\textsc{.link}&int&Item-entity linkage data\\ 
\textsc{.net}&all types&Social graph data\\ \hline
\end{tabular}          
\end{threeparttable} 
\label{tb-atom}
\vspace{-0.cm}
\end{table}

We identify different files by their suffixes.
By summarizing existing recommendation models and datasets, we conclude with four basic data types, i.e., ``\textsf{token}'' (representing integers or strings), ``\textsf{token sequence}'', ``\textsf{float}'' and ``\textsf{float sequence}''.
``\textsf{token}'' and ``\textsf{token sequence}'' are used to represent discrete features such as ID or category, while ``\textsf{float}'' and ``\textsf{float sequence}'' are used to represent  continuous features, such as price.
Atomic files support sparse feature representations, so that the space taken by the atomic files can be largely reduced.  
Most of atomic files support all the four data types except the \textsc{.kg} and \textsc{.link} files. 
Next, we present the detailed description of each atomic file: 

$\bullet$ \textsc{.inter} is a mandatory file  used in all the recommendation tasks. Each line is composed of the user ID (\textsf{token}), item ID (\textsf{token}), user-item rating (\textsf{float}, optional), timestamp (\textsf{float}, optional) and review text (\textsf{token sequence}, optional). Different fields are separated by  commas. 

$\bullet$ \textsc{.user} is a user profile file, which includes users' categorical or continuous features. Each line is formatted as user ID (\textsf{token}), feature (\textsf{token} or \textsf{float}), feature (\textsf{token} or \textsf{float}), ..., feature (\textsf{token} or \textsf{float}).

$\bullet$ \textsc{.item} is an item feature file, which describes the item characteristics, and the format is as follows: item ID (\textsf{token}), feature (\textsf{token} or float), feature (\textsf{token} or \textsf{float}), ..., feature (\textsf{token} or \textsf{float}). \textsc{.user} and \textsc{.item} are used for context-aware recommendation.

$\bullet$ \textsc{.kg} is a knowledge graph file  used for knowledge-based recommendation. Each line corresponds to a $\langle head, tail, relation \rangle$ triplet, and the format is as follows: head entity ID (\textsf{token}), tail entity ID (\textsf{token}), relation ID (\textsf{token}).

$\bullet$ \textsc{.link} is also used for knowledge-based recommendation. It records the correspondence between the recommender systems items and the knowledge graph entities. The file format is as follows: item ID (\textsf{token}), entity ID (\textsf{token}), which denotes the item-to-entity mapping.

$\bullet$ \textsc{.net} is a social network file  used for social recommendation. The format is as follows: source user ID (\textsf{token}), target user ID (\textsf{token}), weight (\textsf{float}, optional).

The essence of the atomic files is feature-based data frames corresponding to different parts of the task input. They can cover the input of most mainstream recommendation tasks in the literature. 
In case the atomic files are not sufficient to support new tasks, one can incrementally introduce new atomic files in a flexible way.

\subsubsection{Input Files for Recommendation Tasks}

Based on the above atomic files, we can utilize a series of file combinations to facilitate five mainstream recommendation tasks, namely \emph{general recommendation}, \emph{context-aware recommendation}, \emph{knowledge-based recommendation}, \emph{sequential recommendation} and \emph{social recommendation}.
Currently, we have implemented the supporting mechanisms for the first four kinds of recommendation tasks, while  the code for social recommendation is under development. 
 
The correspondence between  atomic files and  recommendation models are presented in Table~\ref{tb-cor}.
A major merit of our input files is that atomic files themselves are not dependent on  specific tasks. As we can see, given a dataset, the user can reuse the same \textsc{.inter} file (without any modification  on data files) when switching between different recommendation tasks. Our library reads the configuration file and determines what to do with the data files.    

Another note is that Table~\ref{tb-cor} presents the combination of mandatory atomic files in each task. It is also possible to use additional atomic files besides mandatory  files. For example,
for sequential recommendation, we may also need to use context features.  To support this, one can simply extend the original combination to $\langle$\textsc{.inter}, \textsc{.user}, \textsc{.item}$\rangle$ as needed.

\renewcommand\arraystretch{1.2}
\begin{table}[t]
\centering
\caption{Correspondence between the recommendation task and the atomic files.}
\vspace{-0.cm}
\begin{threeparttable} 
\begin{tabular}{p{4.6cm}<{\centering}||p{3.cm}<{\centering}}
\hline
Tasks &Mandatory atomic files\\ \hline\hline
General Recommendation&\textsc{.inter}\\ 
Context-aware Recommendation&\textsc{.inter}, \textsc{.user}, \textsc{.item}\\
Knowledge-based Recommendation&\textsc{.inter}, \textsc{.kg}, \textsc{.link}\\ 
Sequential Recommendation&\textsc{.inter} \\
Social Recommendation&\textsc{.inter}, \textsc{.net} \\
 \hline
\end{tabular}
\end{threeparttable} 
\label{tb-cor}
\vspace{-0.cm}
\end{table}

\renewcommand\arraystretch{1.2}
\begin{table}[t]
\centering
\caption{The functions that class \textsf{Interaction} supports.}
\vspace{-0.cm}
\label{tb-interaction}
\begin{tabular}{p{2.5cm}<{\centering}||p{5.cm}<{\centering}}
\hline
Function & Description\\
\hline
\hline
\textsf{to(device)} & transfer tensors to \textsf{torch.device} \\
\textsf{cpu} & transfer all tensors to CPU \\
\textsf{numpy} & transfer all tensors to \textsf{numpy.Array} \\ 
\textsf{repeat} & repeats along the \textsf{batch\_size} dimension \\
\textsf{repeat\_interleave} & repeat elements of a tensor\\
\textsf{update} & update an object with other \textsf{Interaction}\\
\hline
\end{tabular}
\vspace{-0.cm}
\end{table}

\subsubsection{The Internal Data Structure \textsc{Interaction}}
As discussed in Section 2.1.1, in our library, \textsc{Interaction} is the internal data structural that is fed into the recommendation algorithms. 

In order to make it unified and flexible, it is implemented as a new abstract data type based on \textsf{python.dict}, which is a key-value indexed data structure.
The keys correspond to  \emph{features} from input, which  can be conveniently referenced with feature names when writing the recommendation algorithms; and the values correspond to \emph{tensors} (implemented by \textsf{torch.Tensor}), which  will be used for the update and computation in learning algorithms.  Specially, the value entry for a specific key stores all the corresponding tensor data in a batch or mini-batch.  

With such a data structure, our library provides a friendly interface to implement the recommendation algorithms in a batch-based mode. 
All the details of the transformation from raw input to internal data representations are transparent to the developers. One can implement  different algorithms easily based on unified internal data representation \textsf{Interaction}.
Besides, the value components are implemented based on \textsf{torch.Tensor}. We wrap many  functions of PyTorch to develop a GRU-oriented data structure, which can support batch-based mechanism (e.g., copying a batch of data to GPU). Specially, we summarize the important functions that \textsf{Interaction} supports in Table~\ref{tb-interaction}.

\renewcommand\arraystretch{1.1}
\begin{table*}[t]
\caption{Implemented \nummodel recommender models in RecBole on 4 categories.}\label{models}
\vspace{-0.2cm}
	\footnotesize
	\scalebox{.9}{
	\begin{tabular}{p{1.8cm}<{\centering}|p{4.cm}<{\centering}|p{1.2cm}<{\centering}|p{1.cm}<{\centering}|p{6.7cm}<{\centering}}
		\hline
		\textbf{Category}                                                             & \textbf{Model}                                       & \textbf{Conference} & \textbf{Year} & \textbf{Typical Evaluation Dataset}                                                    \\ \hline
		\multicolumn{1}{c|}{\multirow{24}{*}{\normalsize{\rotatebox{90}{{General Recommendation}}}}}       & popularity                                           & -                   & -             & -                                                                   \\
		\multicolumn{1}{c|}{}                                                         & ItemKNN~\cite{itemKNN}                               & TOIS                & 2004          & ctlg, ccard, ecmrc, EachMovie, MovieLens, skill                     \\
		\multicolumn{1}{c|}{}                                                         & BPR~\cite{BPRMF}                                     & UAI                 & 2009          & Rossmann, Netflix                                                   \\
		\multicolumn{1}{c|}{}                                                         & SLIMElastic~\cite{SLIM}                              & ICDM                & 2011          & ctlg, ccard, ecmrc,Book-Crossing, MoiveLens, Netflix, Yahoo   Music \\ 
		\multicolumn{1}{c|}{}                                                         & FISM~\cite{FISM}                                     & SIGKDD              & 2013          & MovieLens, Netflix, Yahoo Music                                     \\
		\multicolumn{1}{c|}{}                                                         & LINE~\cite{NNRec,LINE}                                     & WWW                 & 2015          & NetWork~(Wikipedia, Flickr, Youtube, DBLP)                              \\
		\multicolumn{1}{c|}{}                                                         & CDAE~\cite{CDAE}                                     & WSDM                & 2016          & MovieLens, Netflix, Yelp                                            \\
		\multicolumn{1}{c|}{}                                                         & NeuMF~\cite{NeuMF}                                   & WWW                 & 2017          & MovieLens, Pinterest                                                \\
		\multicolumn{1}{c|}{}                                                         & ConvNCF~\cite{ConvNCF}                               & IJCAI               & 2017          & Yelp, Gowalla                                                       \\
		\multicolumn{1}{c|}{}                                                         & DMF~\cite{DMF}                                       & IJCAI               & 2017          & MovieLens, Amazon                                                   \\
		\multicolumn{1}{c|}{}                                                         & NNCF~\cite{NNCF}                                     & CIKM                & 2017          & Delicious, MovieLens, Rossmann                                      \\
		\multicolumn{1}{c|}{}                                                         & NAIS~\cite{NAIS}                                     & TKDE                & 2018          & MovieLens, Pinterest                                                \\
		\multicolumn{1}{c|}{}                                                         & SpectralCF~\cite{SpectralCF}                         & RecSys              & 2018          & MovieLens, HetRec, Amazon                                           \\
		\multicolumn{1}{c|}{}                                                         & MultiVAE~\cite{MultiVAE}                             & WWW                 & 2018          & MovieLens, Million Song, Netflix                                    \\
		\multicolumn{1}{c|}{}                                                         & MultiDAE~\cite{MultiVAE}                             & WWW                 & 2018          & MovieLens, Million Song, Netflix                                    \\
		\multicolumn{1}{c|}{}                                                         & GCMC~\cite{GCMC}                                     & SIGKDD              & 2018          & MovieLens, Flixster,Douban, Yahoo Music                             \\
		\multicolumn{1}{c|}{}                                                         & NGCF~\cite{NGCF}                                     & SIGIR               & 2019          & Gowalla, Yelp, Amazon                                               \\
		\multicolumn{1}{c|}{}                                                         & MacridVAE~\cite{MacridVAE}                           & NeurIPS             & 2019          & AliShop-7C, MovieLens, Netflix                                      \\
		\multicolumn{1}{c|}{}                                                         & EASE~\cite{EASE}                                     & WWW                 & 2019          & MovieLens, Million Song, Netflix                                    \\
		\multicolumn{1}{c|}{}                                                         & LightGCN~\cite{LightGCN}                             & SIGIR               & 2020          & Gowalla, Yelp, Amazon                                               \\
		\multicolumn{1}{c|}{}                                                         & DGCF~\cite{DGCF}                                     & SIGIR               & 2020          & Gowalla, Yelp, Amazon                                               \\
		\multicolumn{1}{c|}{}                                                         & RaCT~\cite{RaCT}                                     & ICLR                & 2020          & MovieLens, Million Song, Netflix                                    \\
		\multicolumn{1}{c|}{}                                                         & RecVAE~\cite{RecVAE}                                 & WSDM                & 2020          & MovieLens, Million Song, Netflix                                    \\
		\multicolumn{1}{c|}{}                                                         & ENMF~\cite{ENMF}                                     & TOIS                & 2020          & Ciao, Epinions, MovieLens                                           \\
		\hline
		\multicolumn{1}{c|}{\multirow{17}{*}{\normalsize{\rotatebox{90}{{Context-aware recommendation}}}}}  & LR~\cite{LR}                                         & WWW                 & 2007          & Microsoft web search dataset                      \\
		\multicolumn{1}{c|}{}                                                         & FM~\cite{FM}                                         & ICDM                & 2010          & CML/PKDD Discovery Challenge 2009, Netflix                          \\
		\multicolumn{1}{c|}{}                                                         & DSSM~\cite{DSSM}                                     & CIKM                & 2013          & Henceforth                                                          \\
		\multicolumn{1}{c|}{}                                                         & FFM~\cite{FFM}                                       & RecSys              & 2016          & Criteo, Avazu                                                       \\
		\multicolumn{1}{c|}{}                                                         & FNN~(DNN)~\cite{FNN}                                  & ECIR                & 2016          & iPinYou                                                             \\
		\multicolumn{1}{c|}{}                                                         & PNN~\cite{PNN}                                       & ICDM                & 2016          & Criteo, iPinYou                                                     \\
		\multicolumn{1}{c|}{}                                                         & Wide\&Deep~\cite{Wide}                               & RecSys              & 2016          & Google play dataset                                       \\
		\multicolumn{1}{c|}{}                                                         & XGBoost~\cite{XGB}                                   & KDD                 & 2016          & Allstate, Higgs Boson, Yahoo LTRC, Criteo                           \\
		\multicolumn{1}{c|}{}                                                         & NFM~\cite{NFM}                                       & SIGIR               & 2017          & Frappe,MovieLens                                                    \\
		\multicolumn{1}{c|}{}                                                         & DeepFM~\cite{DeepFM}                                 & IJCAI               & 2017          & Criteo, Company                                                     \\
		\multicolumn{1}{c|}{}                                                         & AFM~\cite{AFM}                                       & IJCAI               & 2017          & Frappe, MoiveLens                                                   \\
		\multicolumn{1}{c|}{}                                                         & DCN~\cite{DCN}                                       & ADKDD               & 2017          & Criteo                                                              \\
		\multicolumn{1}{c|}{}                                                         & LightGBM~\cite{LightGBM}                             & NIPS                & 2017          & Allstate, Flight Delay, LETOR, KDD10, KDD12                         \\ 
		\multicolumn{1}{c|}{}                                                         & xDeepFM~\cite{xDeepFM}                               & SIGKDD              & 2018          & Criteo, Dianping, Bing News                                         \\
		\multicolumn{1}{c|}{}                                                         & FwFM~\cite{FwFM}                                     & WWW                 & 2018          & Criteo, Oath                                                        \\
		\multicolumn{1}{c|}{}                                                         & DIN~\cite{DIN}                                       & SIGKDD              & 2018          & Amazon, MovieLens, Alibaba                                          \\
			\multicolumn{1}{c|}{}
		&DIEN~\cite{DIEN}                                     & AAAI                &
		2019		  & Amazon                                                               \\
		\multicolumn{1}{c|}{}                                                         & AutoInt~\cite{AutoInt}                               & CIKM                & 2019          & Criteo, Avazu, KDD Cup 2012, MovieLens                              \\
		\hline
		\multicolumn{1}{c|}{\multirow{23}{*}{\normalsize{\rotatebox{90}{{Sequential recommendation}}}}}     & FPMC~\cite{FPMC}                                     & WWW                 & 2010          & ROSSMANN                                                            \\

		\multicolumn{1}{c|}{}                                                         & HRM~\cite{HRM}                                       & SIGIR               & 2015          & Ta-Feng, BeiRen, Tmall                                              \\
		\multicolumn{1}{c|}{}                                                         & Improved GRU-Rec~\cite{ImprovedGRU4Rec}              & DLRS                & 2016          & YOOCHOOSE                                                           \\
		\multicolumn{1}{c|}{}                                                         & GRU4RecF(+feature embedding)~\cite{GRU4RecF}         & RecSys              & 2016          & coined VIDXL, CLASS                                                 \\
		\multicolumn{1}{c|}{}                                                         & Fossil~\cite{Fossil}                                 & ICDM                & 2016          & Amazon, Epinions, Foursquare                                        \\
		\multicolumn{1}{c|}{}                                                         & NARM~\cite{NARM}                                     & CIKM                & 2017          & YOOCHOOSE, DIGINETICA                                               \\
		\multicolumn{1}{c|}{}                                                         & TransRec~\cite{TransRec}                             & RecSys              & 2017          & Amazon, Epinions, Foursquare, Google Local                          \\
		\multicolumn{1}{c|}{}                                                         & STAMP~\cite{STAMP}                                   & SIGKDD              & 2018          & YOOCHOOSE, DIGINETICA                                               \\
		\multicolumn{1}{c|}{}                                                         & Caser~\cite{Caser}                                   & WSDM                & 2018          & MovieLens, Gowalla, Foursquare, Tmall                               \\
		\multicolumn{1}{c|}{}                                                         & SASRec~\cite{SASRec}                                 & ICDM                & 2018          & Amazon, Steam, MovieLens                                            \\
		\multicolumn{1}{c|}{}                                                         & KSR~\cite{KSR}                                       & SIGIR               & 2018          & LastFM, MovieLens, Amazon                                           \\
		\multicolumn{1}{c|}{}                                                         & SHAN~\cite{SHAN}                                     & IJCAI               & 2018          & Tmall, Gowalla                                                      \\
		\multicolumn{1}{c|}{}                                                         & NPE~\cite{NPE}                                       & IJCAI               & 2018          & Movielens, Online Retail, TasteProfile                              \\
		\multicolumn{1}{c|}{}                                                         & NextItnet~\cite{NextItNet}                           & WSDM                & 2019          & YOOCHOOSE, LastFM                                                   \\
		\multicolumn{1}{c|}{}                                                         & BERT4Rec~\cite{BERT4Rec}                             & CIKM                & 2019          & Amazon, Steam, MovieLens                                            \\
		\multicolumn{1}{c|}{}                                                         & SRGNN~\cite{SRGNN}                                   & AAAI                & 2019          & YOOCHOOSE, DIGINETICA                                               \\
		\multicolumn{1}{c|}{}                                                         & GCSAN~\cite{GCSAN}                                   & IJCAI               & 2019          & DIGINETICA, Retailrocket                                            \\
		\multicolumn{1}{c|}{}                                                         & SASRecF(+feature embedding)~\cite{SASRecF}           & IJCAI               & 2019          & -                                                                   \\
		\multicolumn{1}{c|}{}                                                         & FDSA~\cite{FDSA}                                     & IJCAI               & 2019          & Amazon, Tmall                                                       \\
		\multicolumn{1}{c|}{}                                                         & RepeatNet~\cite{RepeatNet}                           & AAAI                & 2019          & YOOCHOOSE, DIGINETICA, LastFM                                       \\
		\multicolumn{1}{c|}{}                                                         & HGN~\cite{HGN}                                       & SIGKDD              & 2019          & MovieLens, Amazon, Goodreads                                        \\ 
		\multicolumn{1}{c|}{}                                                         & S3Rec~\cite{S3Rec}                                   & CIKM                & 2020          & Meituan, Amazon, Yelp, LastFM                                       \\
		\multicolumn{1}{c|}{}                                                         & GRU+KG Embedding                                     & -                   & -             & -                                                                   \\   \hline
		\multicolumn{1}{c|}{\multirow{8}{*}{\normalsize{\rotatebox{90}{\tabincell{c}{Knowledge-based \\ recommendation}}}}} & CKE~\cite{CKE}                                       & SIGKDD              & 2016          & MovieLens, IntentBooks                                              \\
		\multicolumn{1}{c|}{}                                                         & CFKG~\cite{CFKG}                                     & MDPI                & 2018          & Amazon                                                              \\
		\multicolumn{1}{c|}{}                                                         & RippleNet~\cite{RippleNet}                           & CIKM                & 2018          & MovieLens, Book-Crossing, Bing-News                                 \\
		\multicolumn{1}{c|}{}                                                         & KTUP~\cite{KTUP}                                     & WWW                 & 2019          & MovieLens, DBbook2014                                               \\
		\multicolumn{1}{c|}{}                                                         & KGAT~\cite{KGAT}                                     & SIGKDD              & 2019          & Amazon, LastFM, Yelp2018                                            \\
		\multicolumn{1}{c|}{}                                                         & MKR~\cite{MKR}                                       & WWW                 & 2019          & MovieLens, Book-Crossing, LastFM, Bing-News                         \\
		\multicolumn{1}{c|}{}                                                         & KGCN~\cite{KGCN}                                     & WWW                 & 2019          & MovieLens, Book-Crossing, LastFM                                    \\
		\multicolumn{1}{c|}{}                                                         & KGNN-LS~\cite{KGNN-LS}                               & SIGKDD              & 2019          & MovieLens, Book-Crossing, LastFM, Dianping-Food                      \\ \hline
	\end{tabular}
}
\vspace{-0.cm}
\end{table*}

\renewcommand\arraystretch{1.2}
\begin{table}[t]
\caption{Example evaluation settings.}\label{tb-protocol}
\vspace{-0.cm}
\centering
\begin{tabular}{p{1.3cm}<{\centering}||p{6.4cm}<{\centering}}
\hline
{Notation} & {Explanation} \\ 
\hline
\hline
RO\_RS & Random  Ordering + Ratio-based Splitting\\
TO\_LS & Temporal  Ordering + Leave-one-out Splitting\\
RO\_LS & Random  Ordering + Leave-one-out Splitting\\
TO\_RS & Temporal  Ordering + Ratio-based Splitting\\
full & Full ranking with all item candidates\\
uni$N$ & One positive item is paired with $N$ negative items\\
\hline
\end{tabular}
\vspace{-0.cm}
\end{table}

\subsection{Model Module}

Based on the data module, we organize the implementations of recommendation algorithms in a separate model module.  

\subsubsection{Unified Implementation Interface}
By setting up the model module, we can largely decouple the algorithm implementation from other components, which is particularly important to collaborative  development of this library. 
To implement a new model within the four tasks in Table~\ref{tb-cor}, one only needs 
to follow the required interfaces to connect with input and evaluation modules, while the details of other parts can be ignored. 
In specific, we utilize the interface function of \textsf{calculate\_loss}$(\cdot)$ for training and the 
interface function of \textsf{predict}$(\cdot)$ for testing. To implement a model, what a user needs to do is to implement these important interface functions, without considering other details.
These  interface functions are indeed general to various recommendation algorithms, so that we can implement various algorithms in a highly unified way. Such a design mode enables  quick development of new algorithms. 
Besides, our model module further encapsulates many important model implementation details, such as the learning strategy. For code reuse, we implement several  commonly used loss functions (e.g., BPR loss, margin-based loss, and regularization-based loss), neural components (e.g., MLP, multi-head attention, and graph neural network)  and initialization methods (e.g., Xavier's normal and uniform initialization) as individual components, which can be directly used when building complex models or algorithms.

\subsubsection{Implemented Models}
Until now, we have implemented \nummodel recommendation models in the four categories of 
general recommendation, sequential recommendation, context-aware recommendation and knowledge-based recommendation.
We refer the readers to {Table~\ref{models}} for more details on these models.
When selecting the models to be implemented, we have carefully surveyed the recent literature and selected the commonly used recommendation models and their associated variants (which may not receive high citations) in our library. 
We mainly focus on the recently proposed neural methods, while also keep some classic traditional methods such as ItemKNN and FM.
In the future, more methods will also be incorporated in regular update.
For all the implemented models, we have tested their performance on two or four selected datasets, and invited a code reviewer to examine the correctness of the implementation.

\subsubsection{Rich Auxiliary Functions}
In order to better use the models in our library, we also implement a series of useful functions.
A particularly useful function is  automatic parameter tuning. The user is allowed to provide a parameter set for searching an optimal value leading to the best performance. 
Given a set of parameter values, we can indicate four types of tuning methods, i.e.,  
``\emph{Grid Search}'', ``\emph{Random Search}'', ``\emph{Tree of Parzen Estimators~(TPE)}'' and ``\emph{Adaptive TPE}''. The tuning procedure is implemented based on the library of \textsf{hyperopt}~\citep{hyperopt}.
Besides, we add the functions of model saving and loading to store and reuse the learned models, respectively.
Our library also supports the resume of model learning from a previously stored break point.  
In the training process, one can print and monitor the change of the loss value and apply training tricks such as early-stopping. These tiny tricks largely improve the usage experiences with our library.

\subsection{Evaluation Module}

The function of evaluation module is to implement commonly used evaluation protocols for recommender systems. Since different models can be compared under the same evaluation module, our library is useful to standardize the evaluation  of recommender systems. 

\subsubsection{Evaluation Metrics}
Our library supports both value-based and ranking-based evaluation metrics.
The value-based metrics (for rating prediction) include Root Mean Square Error~(RMSE) and Mean Average Error~(MAE), measuring the prediction difference between the true and predicted values.
The ranking-based metrics (for top-$K$ item recommendation) include the most widely used ranking-aware metrics, such as Recall$@K$, Precision$@K$, NDCG, and MRR, measuring the ranking performance of the generated recommendation lists by an algorithm.

\begin{figure*}[htbp]
    \begin{minipage}[b]{0.47\linewidth}
    \centering
    \subfigure[Full Ranking]{
        \includegraphics[width=1.25\linewidth]{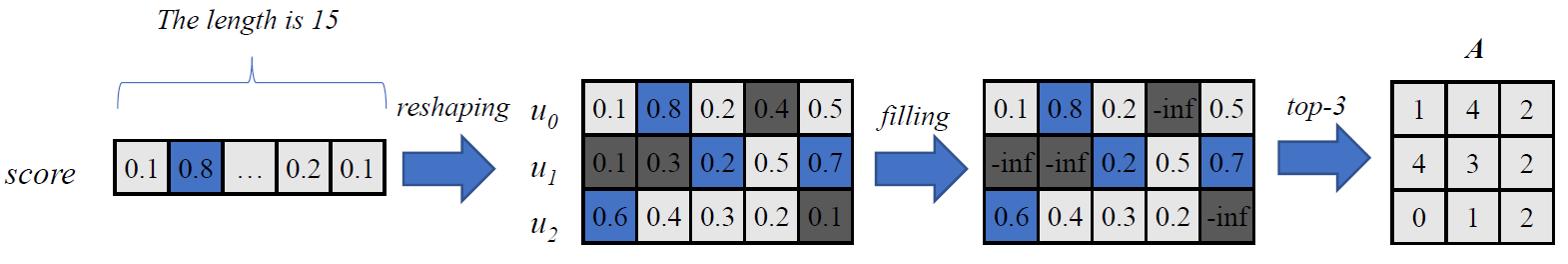}
        \label{fig-full}
    } 
    \subfigure[Sample-based Ranking]{
        \includegraphics[width=1\linewidth]{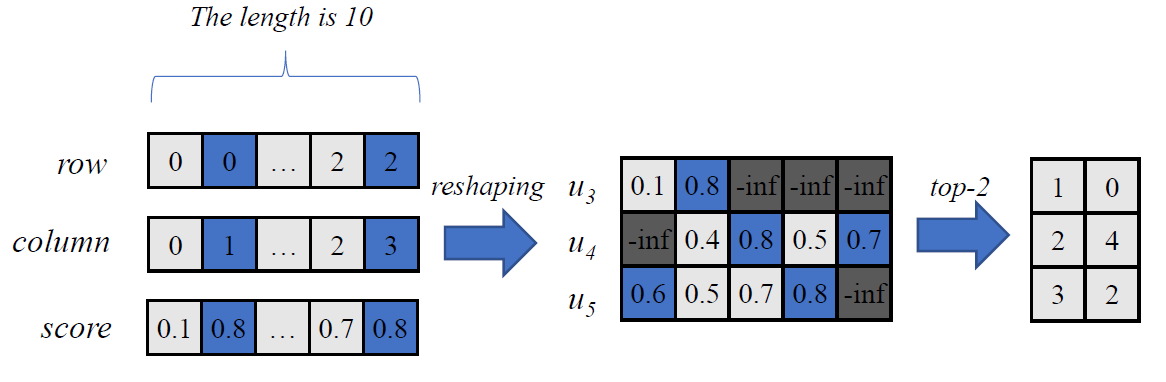}
        \label{fig-sample}
    }
    \end{minipage} 
    \medskip
    \begin{minipage}[b]{0.52\linewidth}
    \centering
    \subfigure[Indexing]{
        \includegraphics[width=0.5\linewidth]{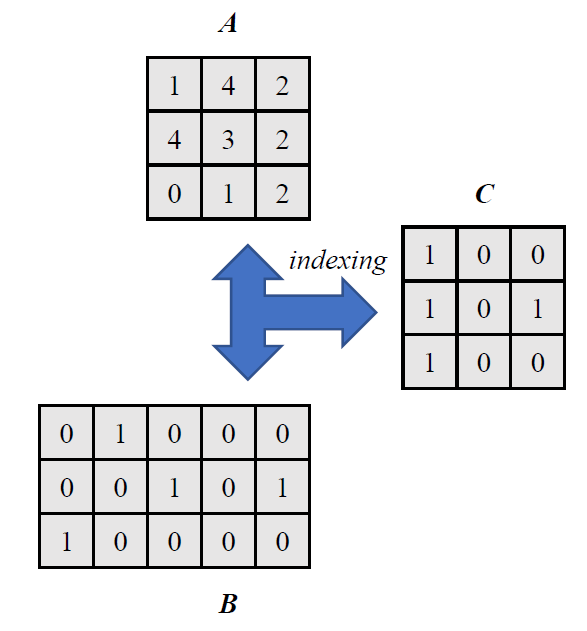}
        \label{fig-index}
    }
    \end{minipage}
    \vspace{-0.cm}
    \caption{Illustration of the proposed acceleration strategy for top-$K$ item evaluation. Here, $u_0,\cdots, u_5$ denote six users; black, blue and grey boxes denote training items, test items and other candidate items, respectively.}
\label{fig-acc}
\end{figure*}

\subsubsection{Evaluation Settings}

In recent years, there are more and more concerns on the appropriate evaluation of recommender systems~\citep{sampledmetrics,comparison}. Basically speaking, the divergence mainly lies in the ranking-based evaluation for top-$K$ item recommendation.   
Note that the focus of our library is not to identify the most suitable evaluation protocols. Instead, we aim to provide most of the widely adopted evaluation protocols (even the most critical ones) in the literature. Our library provides a possibility to compare the performance of various models under different evaluation protocols.

For top-$K$ item recommendation, the implemented evaluation settings cover various settings of our earlier work in \citep{comparison}, where we have studied the influence of different  evaluation protocols on the performance comparison of models. 
In particular, we mainly consider the combinations between item sorting (i.e., how to sort the items before data splitting) and data splitting (i.e., how to derive the train/validation/test sets) for constructing evaluation sets. We also consider both full ranking and sampling-based ranking, which is recently a controversial issue in the field of recommender system~\citep{sampledmetrics}. We summarize the supporting evaluation settings by our library in Table~\ref{tb-protocol}.

In order to facilitate  various evaluation settings, we encapsulate the related functions into four major parts, namely \emph{Group}, \emph{Split}, \emph{Order} and \emph{NegSample}.
With these  implementations, we can effectively support different evaluation protocols, which is also an appealing feature to use our library.  

\subsubsection{Acceleration Strategy for Top-$K$ Evaluation}
Computing Top-$K$ evaluation metrics is usually time consuming.
The basic reason lies in that one need to exhaustively estimate the score for each user-item pair.
Since the method of score estimation varies across different models, it is not easy to optimize the entire evaluation procedure in a general way. 
Therefore, we mainly focus on the step of selecting and generating top $K$ items given the ranking scores.

A problem is that different users have a varying number of ground-truth items in test set (resulting in different-sized user-by-item matrices), which is not suitable for parallel GPU computation in a unified manner. 
Our approach is to consider all the items, including the ones in the training set (called \emph{training items}).
Given $n$ users and $m$ items for consideration, when performing full ranking, we can obtain a $n \times m$ matrix $\bm{D}$ consisting of the confidence scores from a model over the entire item set. When performing sample-based ranking, we create an $n \times m$ matrix $\bm{D}$, initializing all elements to negative infinity. Then, we fill matrix $\bm{D}$ with the confidence scores over sampled items. This step is called \emph{reshaping}.
When performing full ranking with all item candidates, we provide an option to mask the score of training items. If the user choose to mask, the matrix $\bm{D}$ obtained in the above step cannot be directly used for top-$K$ prediction. Our solution is to set the scores of training items to negative infinity, and perform the full ranking over the entire item set without removing training items. This step is called \emph{filling}.
In this way, all the users correspond to equal-sized evaluation matrices (i.e., $n \times m$) for subsequent computation in full ranking and sample-based ranking and the following steps are the same for both cases. 

Then, we utilize the GPU-version \textsf{topk()} function provided by PyTorch to find the top $K$ items with the highest scores for users. The GPU-version \textsf{topk()} function has been specially optimized based on CUDA, which is very efficient in our case. This step is called \emph{topk-finding}.
With the \textsf{topk()} function, we can obtain a matrix $\bm{A}$ with size $n \times K$, which records the original index of the selected top $K$ items. 
We further generate a binary matrix $\bm{B}$ of size $n \times m$ to indicate the existence of an item in the test set (blue boxes in Figure ~\ref{fig-full}) and Figure ~\ref{fig-sample}). Next, we use each row of matrix $\bm{A}$ to index the same row in matrix $\bm{B}$ and obtain a binary matrix $\bm{C}$ of size $n \times K$, which can be implemented efficiently through \textsf{gather()} function provided by PyTorch. We take the case of full ranking as an example in Figure ~\ref{fig-index}. This step is called \emph{indexing}.
Finally, we concatenate the matrix $\bm{C}$ of all the batches. The generated result consists of zeros and ones, which is particularly convenient for computing evaluation metrics. As will be shown next, such an acceleration strategy is able to improve the efficiency for both full ranking and sample-based ranking item recommendation.

\begin{figure}[t]
\centering
\setlength{\fboxrule}{0.pt}
\setlength{\fboxsep}{0.pt}
\fbox{
\includegraphics[width=1.\linewidth]{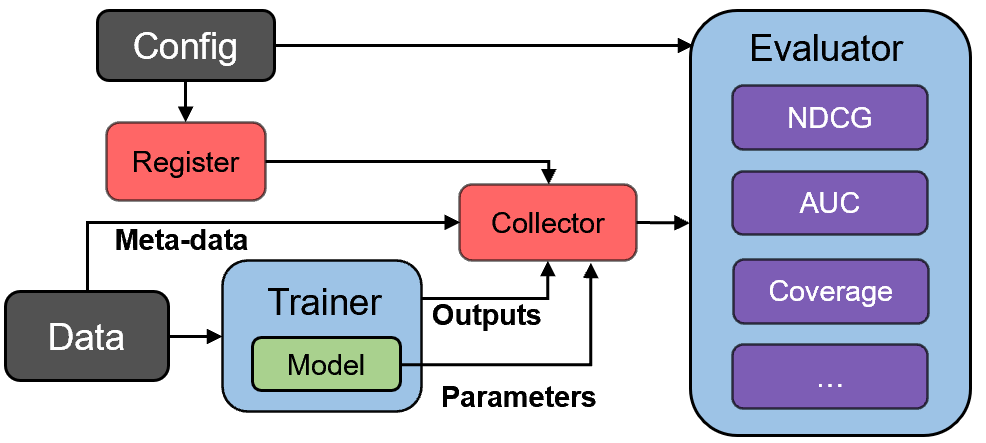}
}
\caption{Architecture and dataflow of the evaluation model.}
\label{fig-evaluator-flow}
\end{figure}

\subsubsection{Efficiency and Scalability}
In this part, we empirically analyze the efficiency improvement yielded by our acceleration strategy and the scalability of the evaluation architecture.
Specifically,  the classic BPR model~\citep{BPRMF} is selected for efficiency analysis, since it is one of the most commonly used baselines for top-$K$ recommendation.
Besides, its model architecture is pretty simple without the influence of other factors, which is suitable for efficiency analysis.   
 We compare its performance  \emph{with} and \emph{without} the acceleration strategy in our implementation.  
We measure the model performance by the total time that (1) it generates a recommendation list of top ten items for users and (2) computes the metrics (NDCG@10 and Recall@10) over the recommendation list, on all the users. 
To further analyze the model efficiency on  datasets of varying sizes, we use three Movielens datasets~\footnote{https://grouplens.org/datasets/movielens/} (i.e., Movielens-100k, Movielens-1M, and MovieLens-10M) to conduct the experiments. We split one original dataset into train, validation and test sets with a ratio of $8:1:1$. We only count the time for generating top ten recommendations (with full ranking) on the test set.  
We average the time of ten runs of different implementations. 
Our experiments are performed on a linux PC with CPU (Intel(R) Xeon(R) Silver 4216, 16 cores, 32 threads, 2.10GHz) and GPU (Nvidia RTX 3090 24G).
The results of efficiency comparison are shown in Table~\ref{t:timecost}.
From the result we can see that by applying the acceleration strategy, we can significantly speed up the evaluation process.
In particular, on the largest dataset MovieLens-10M, the accelerated model can perform the full ranking about two seconds, which indicates that   our implementation is rather efficient. 
Currently, we only compare the entire  time with all the acceleration techniques. As future work, we will analyze the contribution of each specific technique in detail. 
Apart from the superiority of efficiency, the evaluation of \emph{Recbole} is also flexible and extendable. As shown in Figure~\ref{fig-evaluator-flow}, the \emph{evaluator} is decoupled with model and data, and all the required resources for calculating metrics are well-wrapped by a \emph{collector}. 
In this way, it is flexible to develop other customized metrics with these unified interfaces: implement new metrics and sign them in the \emph{register} (see Figure~\ref{fig-evaluator-flow}).

\renewcommand\arraystretch{1.2}
\begin{table}[t]
\caption{Time cost comparison (in second) on different-sized of Movielens datasets with the acceleration strategy or not. BPR$^{acc}$ denotes the model with acceleration strategy.}\label{t:timecost}
\centering
\begin{tabular}{c||c|c|c}
\hline
\multirow{2}{*}{Model} & \multicolumn{3}{c}{Dataset} \\ \cline{2-4} 
& MovieLens-100k & MovieLens-1M & MovieLens-10M \\ \hline\hline
BPR & 0.245s & 2.478s & 29.900s \\ \hline
BPR$^{acc}$ & 0.009s & 0.090s & 2.210s \\ \hline
\end{tabular}
\end{table}

\section{Usage Examples of the Library}
In this section, we show how to use our library with code examples.
We detail the usage description in two parts, namely running existing models in our library and implementing  new models based on the interfaces provided in our library.

\begin{figure}[t]
\centering
\setlength{\fboxrule}{0.pt}
\setlength{\fboxsep}{0.pt}
\fbox{
\includegraphics[width=1.\linewidth]{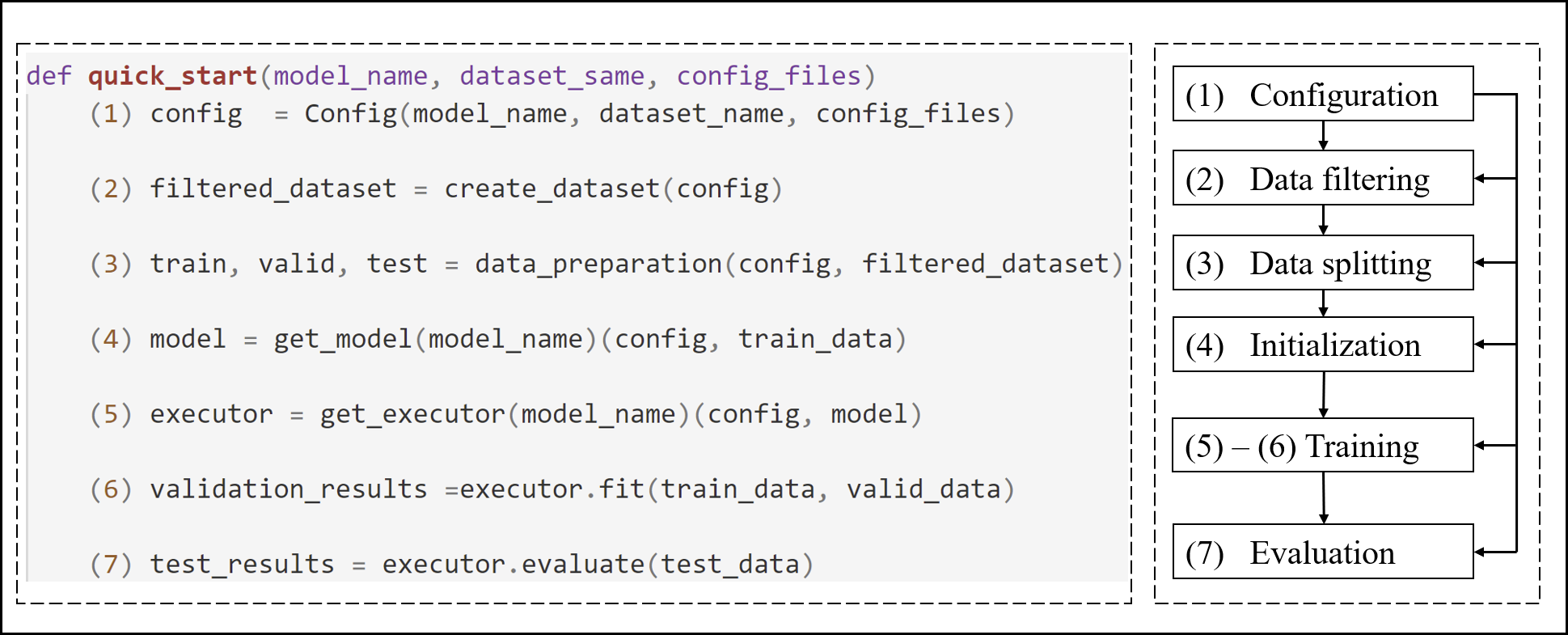}
}
\caption{An illustrative usage flow of our library.
}
\vspace{-0.cm}
\label{usage}
\end{figure}

\subsection{Running Existing Models}
The contained models in our library can be run with either fixed parameters or auto-tuned parameters.

\subsubsection{Model Running with Fixed Parameters}
Figure~\ref{usage} presents a general procedure for running existing models in our library.
To begin with, one needs to download and format the raw public dataset based on our provided utils.
The running procedure relies on some experimental configuration, which can be obtained from the files, command line or parameter dictionaries.
The dataset and model are prepared according to the configured parameters and settings, and the execution module is responsible for training and evaluating the models.
The detailed steps are given as follows.

(\romannumeral1) Formatting the dataset.
A dataset is firstly selected by a user, and then it is formatted based on the scripts, which can generate required atomic files for different datasets.
This procedure takes the following code: $\textsf{\text{atomic\_file}=\text{PreProcess}(\text{dataset})}$.

(\romannumeral2) Generating the configuration.
In our library, the experiment configurations can be generated in different ways.
One can write a configuration file, and then read this file in the main function as in line (1) of Figure~\ref{usage}.
Another way for configuration is to include parameters in the command line, which is useful for specially focused parameters.
At last, one can also directly write the parameter dictionaries in the code.

(\romannumeral3) Filtering and splitting the dataset. We provide rich 
auxiliary functions to filter and split the datasets.
For example, one can filter the dataset by keeping the users/items with at least $K$ interactions, removing the data occurred in some fixed time period.  
Different filtering methods can be applied in a unified function (line (2) of Figure~\ref{usage}). 
When splitting the dataset, one can indicate ratio-based method or leave-one-out method. Then, one can use line (3) of Figure~\ref{usage} to generate the training, validation and testing sets.

(\romannumeral4)  Loading the model.
The next step is to build a recommender model.
Given a target model in mind, the user can obtain a model instance according to line (4) of Figure~\ref{usage}.

(\romannumeral5) Training and evaluation. 
Once the dataset and model are prepared, the user can, at last, train and evaluate the model based on line (5) of Figure~\ref{usage}.

\renewcommand\arraystretch{1.2}
\begin{table*}[t]
\caption{Comparison with existing recommender system libraries.}
\vspace{-0.cm}
\centering
\setlength{\tabcolsep}{3.pt}
\begin{threeparttable}  
\begin{tabular}
{p{4cm}<{\centering}||p{1.6cm}<{\centering}p{1.2cm}<{\centering}p{1.2cm}<{\centering}p{1.1cm}<{\centering}p{1.1cm}<{\centering}p{1.1cm}<{\centering}p{1.6cm}<{\centering}p{1.cm}<{\centering}p{1.4cm}<{\centering}}\hline
Library&Language&\#Models&\#Datasets&\#Fork&\#Star&\#Issues&Release time&Neural&PT\\\hline\hline
MyMediaLite~(\cite{MyMediaLite})&C\#&61&5&199&477&451&2010&No&manual\\\hline
LibFM~(\cite{LibFM})&C++&1&-&415&1400&32&2014&No&manual\\\hline
LibRec~(\cite{Librec})&Java&93&11&1000+&3009&252&2014&No&manual\\\hline
RankSys~(\cite{RankSys})&Java&8&-&58&259&38&2016&No&manual\\\hline
Crab~(\cite{Crab})&Python&2&4&381&1122&75&2011&No&manual\\\hline
Surprise~(\cite{Surprise})&Python&11&3&888&4989&333&2015&No&manual\\\hline
LightFM~(\cite{LightFM})&Python&1&2&610&3741&425&2015&No&manual\\\hline
Case Recommender~(\cite{CaseRecommender})&Python&27&-&75&354&24&2015&No&manual\\\hline
Recommenders~(\cite{Recommenders})&Tensorflow&31&5&1900+&10000+&602&2018&Yes&automatic\\\hline
Cornac~(\cite{Cornac})&Tensorflow&42&14&75&397&58&2018&Yes&automatic\\\hline
NeuRec~(\cite{NeuRec})&Tensorflow&33&3&199&816&29&2019&Yes&manual\\\hline
Elliot~(\cite{Elliot})&Tensorflow&50&-&19&108&8&2021&Yes&automatic\\\hline
Spotlight~(\cite{Spotlight})&PyTorch&8&5&389&2552&109&2017&Yes&automatic\\\hline
DaisyRec~(\cite{daisyRec})&PyTorch&20&14&59&354&8&2019&Yes&automatic\\\hline
ReChorus~(\cite{ReChorus})&PyTorch&12&2&47&214&16&2020&Yes&manual\\\hline
Beta-recsys~(\cite{Beta-recsys})&PyTorch&22&21&25&75&120&2020&Yes&manual\\\hline\hline
\textbf{\modelname}&\textbf{PyTorch}&\textbf{\nummodel}&\textbf{\numdataset}&\textbf{193}&\textbf{1179}&\textbf{163}&\textbf{2020}&\textbf{Yes}&\textbf{automatic}\\\hline
\end{tabular}
\begin{tablenotes}    
   \footnotesize              
   \item[1] Neural means the libarary support deep recommender models, and PT denotes parameter tuning.  The statistics were collected on the date of Aug 21, 2021.
\end{tablenotes}            
\end{threeparttable}   
\vspace{-0.cm} 
\label{tab:compare}  
\end{table*}

\subsubsection{Parameter Tuning}
Our library is featured in the capability of automatic parameter (or hyper-parameter) tuning.
One can readily optimize a given model according to the provided hyper-parameter range.
The general steps are given as follows.

(\romannumeral1) Setting the parameter range. 
The users are allowed to provide candidate parameter values in the file ``hyper.test''.
In this file, each line is formatted as: $\textsf{\text{parameter}=[\text{value $1$},\text{value $2$},...\text{value $n$}]}$.
Instead of a fixed value, the users can empirically indicate a value set, which will be explored in the following tuning steps. 

(\romannumeral2) Setting the tuning method. 
Our parameter tuning function is implemented based on the library \textsf{hyperopt}.
Given a set of parameter values, we can indicate four types of tuning methods, i.e., 
``\emph{Grid Search}'', ``\emph{Random Search}'', ``\emph{Tree of Parzen Estimators~(TPE)}'' and ``\emph{Adaptive TPE}''.
The tuning method is invoked by the following code: $\textsf{\text{hy} = \text{HyperTuning}(\text{objective},\ \text{tuning\_method},\ \text{range\_file})}$,
where the parameter range file is used to indicate parameter values.

(\romannumeral3) Starting the tuning process.
The user can start the running process by the following code: $\textsf{\text{hy.run()}}$.
With the tuning ranges and method, our library will run the model iteratively, and finally output and save the optimal parameters and the corresponding model performance.

\subsection{Implementing a New Model}
Based on RecBole, it is convenient to implement a new model by instantiating three functions as follows:

(\romannumeral1) Implementing the ``\textsf{\_\_init\_\_()}'' function.
In this function, the user performs parameter initialization, global variable definition and so on.
The new model should be a sub-class of the abstract model class provided in our library.
Until now, we have implemented the abstract classes for  general recommendation, knowledge-based recommendation, sequential recommendation and context-aware recommendation.

(\romannumeral2) Implementing the ``\textsf{calculate\_loss()}'' function.
This function calculates the loss to be  optimized by the new model.
Based on the return value of this function, the library will automatically invoke different optimization methods to learn the model according to the pre-set configurations.

(\romannumeral3) Implementing the ``\textsf{predict()}'' function.
This function is used to predict a score from the input data (e.g., the rating given a user-item pair).
This function can be used to compute the loss or derive the item ranking during the model testing phase.

\section{Comparison with Existing Libraries}
In recent years, a considerable number of open source recommender system libraries have been released for research purpose. We summarize and compare the characteristics of existing recommender system libraries in Table~\ref{tab:compare}, from which, we can see: the programming language of these libraries gradually evolves from C/C++/JAVA to Python/Tensorflow/PyTorch. From the model perspective, recent libraries mostly support neural recommender models, which agrees with the advance trend in the recommendation domain. In our framework, we select PyTorch as the basic deep learning framework  for development, due to its friendly features like easy debugging, compatible with numpy and etc.

RecBole provides the most comprehensive models and benchmark datasets among existing libraries, which can better free the users from the heavy model re-programming work. 
In addition to reproduce the existing models, we aim to ease the developing process of new algorithms. We design general and extensible underlying data structures to support the unified development framework. By providing a series of useful tools, functions and scripts (e.g., automatic parameter tuning), our library is particularly convenient to be used for scientific research. 

At last, we believe that implementation is only the first step for open source recommendation library, as more efforts are needed to maintain and update the library according to users' feedbacks and suggestions.
Our team is working hard to respond to the GitHub issues and fix possible bugs (134 issues were solved until Aug 21, 2021).  
After release, our library has received much attention from the users. To the date of publication, it is ranked at the third and fourth places based on the number of received stars for the topic of ``recommender system'' and ``recommendation system'', respectively.

\ignore{At last, we believe implementing the library is only the first step towards providing the recommendation researchers with a better experiment environment.
Much more efforts are needed to maintain this library, and receive suggestions from the users.
To this end, we keep on fixing bugs and solving issues since the first release of our library.
Until now, we have solved nearly 130 issues, and obtained 1049 stars and 160 forks in only 8 months.
Because of our effort in the past few months, our library is ranked at 5 on Github under the ``recommendation'' and `recommendation system'' queries, respectively, which is a positive response on our hard work and confirms the popularity and acceptability of our library.
In the future, we will actively improve our library to continually serve the recommendation community, and we believe this is not hard for us to promise, since the project leaders are all faculties in the universities, who can easily stick to their research interest in the recommendation domain without too much disturbation.
}

\section{Conclusion}
In this paper, we have released a new recommender system library called \modelname.  
So far, we have implemented \nummodel recommendation algorithms on \numdataset commonly used datasets. 
We design general and extensible data structures to offer a unified development framework for new recommendation algorithms. 
We also support extensive and standard evaluation protocols to compare and test different recommendation algorithms. 
Besides, our library is implemented in a GPU-accelerated way, involving a series of optimization techniques for achieving efficient execution.  
The \modelname library is expected to improve the  reproducibility of recommendation models, ease the developing process of new algorithms,  and set up a benchmark framework for the field of recommender system.
In the future, we will make continuous efforts to  add more  datasets and models.
We will also consider adding more utils for facilitating the usage of our library, such as result visualization and algorithm debugging. 

\section{Acknowledgment}
This work was partially supported by the National Natural Science Foundation of China under Grant No. 61872369, 61802029 and 61972155, Beijing Academy of Artificial Intelligence (BAAI) under Grant No. BAAI2020ZJ0301, and Beijing Outstanding Young Scientist Program under Grant No. BJJWZYJH012019100020098.
We also sincerely thank non-author team members (Chen Yang, Zheng Gong, Chenzhan Shang, Zihan Song, Ze Zhang, Jingsen Zhang, Lanling Xu, Zhen Tian and Wenjing Yue) for testing our library. 

\newpage
\bibliographystyle{ACM-Reference-Format}
\balance
\bibliography{main}


\begin{thebibliography}{92}


\ifx \showCODEN    \undefined \def \showCODEN     #1{\unskip}     \fi
\ifx \showDOI      \undefined \def \showDOI       #1{#1}\fi
\ifx \showISBNx    \undefined \def \showISBNx     #1{\unskip}     \fi
\ifx \showISBNxiii \undefined \def \showISBNxiii  #1{\unskip}     \fi
\ifx \showISSN     \undefined \def \showISSN      #1{\unskip}     \fi
\ifx \showLCCN     \undefined \def \showLCCN      #1{\unskip}     \fi
\ifx \shownote     \undefined \def \shownote      #1{#1}          \fi
\ifx \showarticletitle \undefined \def \showarticletitle #1{#1}   \fi
\ifx \showURL      \undefined \def \showURL       {\relax}        \fi
\providecommand\bibfield[2]{#2}
\providecommand\bibinfo[2]{#2}
\providecommand\natexlab[1]{#1}
\providecommand\showeprint[2][]{arXiv:#2}

\bibitem[\protect\citeauthoryear{Ai, Azizi, Chen, and Zhang}{Ai
  et~al\mbox{.}}{2018}]%
        {CFKG}
\bibfield{author}{\bibinfo{person}{Qingyao Ai}, \bibinfo{person}{Vahid Azizi},
  \bibinfo{person}{Xu Chen}, {and} \bibinfo{person}{Yongfeng Zhang}.}
  \bibinfo{year}{2018}\natexlab{}.
\newblock \showarticletitle{Learning Heterogeneous Knowledge Base Embeddings
  for Explainable Recommendation}.
\newblock \bibinfo{journal}{\emph{Algorithms}} \bibinfo{volume}{11},
  \bibinfo{number}{9} (\bibinfo{year}{2018}), \bibinfo{pages}{137}.
\newblock


\bibitem[\protect\citeauthoryear{Anelli, Bellog{\'{\i}}n, Ferrara, Malitesta,
  Merra, Pomo, Donini, and Noia}{Anelli et~al\mbox{.}}{2021}]%
        {Elliot}
\bibfield{author}{\bibinfo{person}{Vito~Walter Anelli},
  \bibinfo{person}{Alejandro Bellog{\'{\i}}n}, \bibinfo{person}{Antonio
  Ferrara}, \bibinfo{person}{Daniele Malitesta},
  \bibinfo{person}{Felice~Antonio Merra}, \bibinfo{person}{Claudio Pomo},
  \bibinfo{person}{Francesco~Maria Donini}, {and} \bibinfo{person}{Tommaso~Di
  Noia}.} \bibinfo{year}{2021}\natexlab{}.
\newblock \showarticletitle{Elliot: {A} Comprehensive and Rigorous Framework
  for Reproducible Recommender Systems Evaluation}. In
  \bibinfo{booktitle}{\emph{{SIGIR}}}. \bibinfo{publisher}{{ACM}},
  \bibinfo{pages}{2405--2414}.
\newblock


\bibitem[\protect\citeauthoryear{Argyriou, Gonz{\'{a}}lez{-}Fierro, and
  Zhang}{Argyriou et~al\mbox{.}}{2020}]%
        {Recommenders}
\bibfield{author}{\bibinfo{person}{Andreas Argyriou}, \bibinfo{person}{Miguel
  Gonz{\'{a}}lez{-}Fierro}, {and} \bibinfo{person}{Le Zhang}.}
  \bibinfo{year}{2020}\natexlab{}.
\newblock \showarticletitle{Microsoft Recommenders: Best Practices for
  Production-Ready Recommendation Systems}. In
  \bibinfo{booktitle}{\emph{Companion of The 2020 Web Conference 2020, Taipei,
  Taiwan, April 20-24, 2020}}. \bibinfo{pages}{50--51}.
\newblock


\bibitem[\protect\citeauthoryear{Bai, Wen, Zhang, and Zhao}{Bai
  et~al\mbox{.}}{2017}]%
        {NNCF}
\bibfield{author}{\bibinfo{person}{Ting Bai}, \bibinfo{person}{Ji{-}Rong Wen},
  \bibinfo{person}{Jun Zhang}, {and} \bibinfo{person}{Wayne~Xin Zhao}.}
  \bibinfo{year}{2017}\natexlab{}.
\newblock \showarticletitle{A Neural Collaborative Filtering Model with
  Interaction-based Neighborhood}. In \bibinfo{booktitle}{\emph{{CIKM}}}.
  \bibinfo{publisher}{{ACM}}, \bibinfo{pages}{1979--1982}.
\newblock


\bibitem[\protect\citeauthoryear{Bergstra, Yamins, and Cox}{Bergstra
  et~al\mbox{.}}{2013}]%
        {hyperopt}
\bibfield{author}{\bibinfo{person}{James Bergstra}, \bibinfo{person}{Daniel
  Yamins}, {and} \bibinfo{person}{David~D. Cox}.}
  \bibinfo{year}{2013}\natexlab{}.
\newblock \showarticletitle{Making a Science of Model Search: Hyperparameter
  Optimization in Hundreds of Dimensions for Vision Architectures}. In
  \bibinfo{booktitle}{\emph{Proceedings of the 30th International Conference on
  Machine Learning, {ICML} 2013, Atlanta, GA, USA, 16-21 June 2013}}
  \emph{(\bibinfo{series}{{JMLR} Workshop and Conference Proceedings},
  Vol.~\bibinfo{volume}{28})}. \bibinfo{publisher}{JMLR.org},
  \bibinfo{pages}{115--123}.
\newblock


\bibitem[\protect\citeauthoryear{Cao, Wang, He, Hu, and Chua}{Cao
  et~al\mbox{.}}{2019}]%
        {KTUP}
\bibfield{author}{\bibinfo{person}{Yixin Cao}, \bibinfo{person}{Xiang Wang},
  \bibinfo{person}{Xiangnan He}, \bibinfo{person}{Zikun Hu}, {and}
  \bibinfo{person}{Tat{-}Seng Chua}.} \bibinfo{year}{2019}\natexlab{}.
\newblock \showarticletitle{Unifying Knowledge Graph Learning and
  Recommendation: Towards a Better Understanding of User Preferences}. In
  \bibinfo{booktitle}{\emph{The World Wide Web Conference, {WWW} 2019, San
  Francisco, CA, USA, May 13-17, 2019}}. \bibinfo{pages}{151--161}.
\newblock


\bibitem[\protect\citeauthoryear{Caraciolo, Melo, and Caspirro}{Caraciolo
  et~al\mbox{.}}{2011}]%
        {Crab}
\bibfield{author}{\bibinfo{person}{Marcel Caraciolo}, \bibinfo{person}{Bruno
  Melo}, {and} \bibinfo{person}{Ricardo Caspirro}.}
  \bibinfo{year}{2011}\natexlab{}.
\newblock \showarticletitle{Crab: A recommendation engine framework for
  python}.
\newblock \bibinfo{journal}{\emph{Jarrodmillman Com}} (\bibinfo{year}{2011}).
\newblock


\bibitem[\protect\citeauthoryear{Castells, Hurley, and Vargas}{Castells
  et~al\mbox{.}}{2015}]%
        {RankSys}
\bibfield{author}{\bibinfo{person}{Pablo Castells}, \bibinfo{person}{Neil~J.
  Hurley}, {and} \bibinfo{person}{Saul Vargas}.}
  \bibinfo{year}{2015}\natexlab{}.
\newblock \showarticletitle{Novelty and Diversity in Recommender Systems}.
\newblock In \bibinfo{booktitle}{\emph{Recommender Systems Handbook}}.
  \bibinfo{pages}{881--918}.
\newblock


\bibitem[\protect\citeauthoryear{Chen, Zhang, Zhang, Liu, and Ma}{Chen
  et~al\mbox{.}}{2020}]%
        {ENMF}
\bibfield{author}{\bibinfo{person}{Chong Chen}, \bibinfo{person}{Min Zhang},
  \bibinfo{person}{Yongfeng Zhang}, \bibinfo{person}{Yiqun Liu}, {and}
  \bibinfo{person}{Shaoping Ma}.} \bibinfo{year}{2020}\natexlab{}.
\newblock \showarticletitle{Efficient Neural Matrix Factorization without
  Sampling for Recommendation}.
\newblock \bibinfo{journal}{\emph{{ACM} Trans. Inf. Syst.}}
  \bibinfo{volume}{38}, \bibinfo{number}{2} (\bibinfo{year}{2020}),
  \bibinfo{pages}{14:1--14:28}.
\newblock


\bibitem[\protect\citeauthoryear{Chen and Guestrin}{Chen and Guestrin}{2016}]%
        {XGB}
\bibfield{author}{\bibinfo{person}{Tianqi Chen} {and} \bibinfo{person}{Carlos
  Guestrin}.} \bibinfo{year}{2016}\natexlab{}.
\newblock \showarticletitle{XGBoost: {A} Scalable Tree Boosting System}. In
  \bibinfo{booktitle}{\emph{{KDD}}}. \bibinfo{publisher}{{ACM}},
  \bibinfo{pages}{785--794}.
\newblock


\bibitem[\protect\citeauthoryear{Cheng, Koc, Harmsen, Shaked, Chandra, Aradhye,
  Anderson, Corrado, Chai, Ispir, Anil, Haque, Hong, Jain, Liu, and Shah}{Cheng
  et~al\mbox{.}}{2016}]%
        {Wide}
\bibfield{author}{\bibinfo{person}{Heng{-}Tze Cheng}, \bibinfo{person}{Levent
  Koc}, \bibinfo{person}{Jeremiah Harmsen}, \bibinfo{person}{Tal Shaked},
  \bibinfo{person}{Tushar Chandra}, \bibinfo{person}{Hrishi Aradhye},
  \bibinfo{person}{Glen Anderson}, \bibinfo{person}{Greg Corrado},
  \bibinfo{person}{Wei Chai}, \bibinfo{person}{Mustafa Ispir},
  \bibinfo{person}{Rohan Anil}, \bibinfo{person}{Zakaria Haque},
  \bibinfo{person}{Lichan Hong}, \bibinfo{person}{Vihan Jain},
  \bibinfo{person}{Xiaobing Liu}, {and} \bibinfo{person}{Hemal Shah}.}
  \bibinfo{year}{2016}\natexlab{}.
\newblock \showarticletitle{Wide {\&} Deep Learning for Recommender Systems}.
  In \bibinfo{booktitle}{\emph{Proceedings of the 1st Workshop on Deep Learning
  for Recommender Systems, DLRS@RecSys 2016, Boston, MA, USA, September 15,
  2016}}. \bibinfo{pages}{7--10}.
\newblock


\bibitem[\protect\citeauthoryear{Costa, Fressato, Neto, Manzato, and
  Campello}{Costa et~al\mbox{.}}{2018}]%
        {CaseRecommender}
\bibfield{author}{\bibinfo{person}{Arthur F.~Da Costa},
  \bibinfo{person}{Eduardo~P. Fressato}, \bibinfo{person}{Fernando S.~Aguiar
  Neto}, \bibinfo{person}{Marcelo~G. Manzato}, {and} \bibinfo{person}{Ricardo
  J. G.~B. Campello}.} \bibinfo{year}{2018}\natexlab{}.
\newblock \showarticletitle{Case recommender: a flexible and extensible python
  framework for recommender systems}. In \bibinfo{booktitle}{\emph{Proceedings
  of the 12th {ACM} Conference on Recommender Systems, RecSys 2018, Vancouver,
  BC, Canada, October 2-7, 2018}}. \bibinfo{pages}{494--495}.
\newblock


\bibitem[\protect\citeauthoryear{Deshpande and Karypis}{Deshpande and
  Karypis}{2004}]%
        {itemKNN}
\bibfield{author}{\bibinfo{person}{Mukund Deshpande} {and}
  \bibinfo{person}{George Karypis}.} \bibinfo{year}{2004}\natexlab{}.
\newblock \showarticletitle{Item-based top-\emph{N} recommendation algorithms}.
\newblock \bibinfo{journal}{\emph{{ACM} Trans. Inf. Syst.}}
  \bibinfo{volume}{22}, \bibinfo{number}{1} (\bibinfo{year}{2004}),
  \bibinfo{pages}{143--177}.
\newblock


\bibitem[\protect\citeauthoryear{Gantner, Rendle, Freudenthaler, and
  Schmidt-Thieme}{Gantner et~al\mbox{.}}{2011}]%
        {MyMediaLite}
\bibfield{author}{\bibinfo{person}{Zeno Gantner}, \bibinfo{person}{Steffen
  Rendle}, \bibinfo{person}{Christoph Freudenthaler}, {and}
  \bibinfo{person}{Lars Schmidt-Thieme}.} \bibinfo{year}{2011}\natexlab{}.
\newblock \showarticletitle{{MyMediaLite}: A Free Recommender System Library}.
  In \bibinfo{booktitle}{\emph{5th ACM International Conference on Recommender
  Systems (RecSys 2011)}} (Chicago, USA).
\newblock


\bibitem[\protect\citeauthoryear{Guo, Zhang, Sun, and Yorke{-}Smith}{Guo
  et~al\mbox{.}}{2015}]%
        {Librec}
\bibfield{author}{\bibinfo{person}{Guibing Guo}, \bibinfo{person}{Jie Zhang},
  \bibinfo{person}{Zhu Sun}, {and} \bibinfo{person}{Neil Yorke{-}Smith}.}
  \bibinfo{year}{2015}\natexlab{}.
\newblock \showarticletitle{LibRec: {A} Java Library for Recommender Systems}.
  In \bibinfo{booktitle}{\emph{Posters, Demos, Late-breaking Results and
  Workshop Proceedings of the 23rd Conference on User Modeling, Adaptation, and
  Personalization {(UMAP} 2015), Dublin, Ireland, June 29 - July 3, 2015}}.
\newblock


\bibitem[\protect\citeauthoryear{Guo, Tang, Ye, Li, and He}{Guo
  et~al\mbox{.}}{2017}]%
        {DeepFM}
\bibfield{author}{\bibinfo{person}{Huifeng Guo}, \bibinfo{person}{Ruiming
  Tang}, \bibinfo{person}{Yunming Ye}, \bibinfo{person}{Zhenguo Li}, {and}
  \bibinfo{person}{Xiuqiang He}.} \bibinfo{year}{2017}\natexlab{}.
\newblock \showarticletitle{DeepFM: {A} Factorization-Machine based Neural
  Network for {CTR} Prediction}. In \bibinfo{booktitle}{\emph{Proceedings of
  the Twenty-Sixth International Joint Conference on Artificial Intelligence,
  {IJCAI} 2017, Melbourne, Australia, August 19-25, 2017}}.
  \bibinfo{pages}{1725--1731}.
\newblock


\bibitem[\protect\citeauthoryear{He, Kang, and McAuley}{He
  et~al\mbox{.}}{2017a}]%
        {TransRec}
\bibfield{author}{\bibinfo{person}{Ruining He}, \bibinfo{person}{Wang{-}Cheng
  Kang}, {and} \bibinfo{person}{Julian~J. McAuley}.}
  \bibinfo{year}{2017}\natexlab{a}.
\newblock \showarticletitle{Translation-based Recommendation}. In
  \bibinfo{booktitle}{\emph{Proceedings of the Eleventh {ACM} Conference on
  Recommender Systems, RecSys 2017, Como, Italy, August 27-31, 2017}}.
  \bibinfo{pages}{161--169}.
\newblock


\bibitem[\protect\citeauthoryear{He and McAuley}{He and McAuley}{2016}]%
        {Fossil}
\bibfield{author}{\bibinfo{person}{Ruining He} {and} \bibinfo{person}{Julian~J.
  McAuley}.} \bibinfo{year}{2016}\natexlab{}.
\newblock \showarticletitle{Fusing Similarity Models with Markov Chains for
  Sparse Sequential Recommendation}. In \bibinfo{booktitle}{\emph{{ICDM}}}.
  \bibinfo{publisher}{{IEEE} Computer Society}, \bibinfo{pages}{191--200}.
\newblock


\bibitem[\protect\citeauthoryear{He and Chua}{He and Chua}{2017}]%
        {NFM}
\bibfield{author}{\bibinfo{person}{Xiangnan He} {and}
  \bibinfo{person}{Tat{-}Seng Chua}.} \bibinfo{year}{2017}\natexlab{}.
\newblock \showarticletitle{Neural Factorization Machines for Sparse Predictive
  Analytics}. In \bibinfo{booktitle}{\emph{Proceedings of the 40th
  International {ACM} {SIGIR} Conference on Research and Development in
  Information Retrieval, Shinjuku, Tokyo, Japan, August 7-11, 2017}}.
  \bibinfo{pages}{355--364}.
\newblock


\bibitem[\protect\citeauthoryear{He, Deng, Wang, Li, Zhang, and Wang}{He
  et~al\mbox{.}}{2020}]%
        {LightGCN}
\bibfield{author}{\bibinfo{person}{Xiangnan He}, \bibinfo{person}{Kuan Deng},
  \bibinfo{person}{Xiang Wang}, \bibinfo{person}{Yan Li},
  \bibinfo{person}{Yong{-}Dong Zhang}, {and} \bibinfo{person}{Meng Wang}.}
  \bibinfo{year}{2020}\natexlab{}.
\newblock \showarticletitle{LightGCN: Simplifying and Powering Graph
  Convolution Network for Recommendation}. In
  \bibinfo{booktitle}{\emph{Proceedings of the 43rd International {ACM} {SIGIR}
  conference on research and development in Information Retrieval, {SIGIR}
  2020, Virtual Event, China, July 25-30, 2020}}. \bibinfo{pages}{639--648}.
\newblock


\bibitem[\protect\citeauthoryear{He, Du, Wang, Tian, Tang, and Chua}{He
  et~al\mbox{.}}{2018a}]%
        {ConvNCF}
\bibfield{author}{\bibinfo{person}{Xiangnan He}, \bibinfo{person}{Xiaoyu Du},
  \bibinfo{person}{Xiang Wang}, \bibinfo{person}{Feng Tian},
  \bibinfo{person}{Jinhui Tang}, {and} \bibinfo{person}{Tat{-}Seng Chua}.}
  \bibinfo{year}{2018}\natexlab{a}.
\newblock \showarticletitle{Outer Product-based Neural Collaborative
  Filtering}. In \bibinfo{booktitle}{\emph{Proceedings of the Twenty-Seventh
  International Joint Conference on Artificial Intelligence, {IJCAI} 2018, July
  13-19, 2018, Stockholm, Sweden}}. \bibinfo{pages}{2227--2233}.
\newblock


\bibitem[\protect\citeauthoryear{He, He, Song, Liu, Jiang, and Chua}{He
  et~al\mbox{.}}{2018b}]%
        {NAIS}
\bibfield{author}{\bibinfo{person}{Xiangnan He}, \bibinfo{person}{Zhankui He},
  \bibinfo{person}{Jingkuan Song}, \bibinfo{person}{Zhenguang Liu},
  \bibinfo{person}{Yu{-}Gang Jiang}, {and} \bibinfo{person}{Tat{-}Seng Chua}.}
  \bibinfo{year}{2018}\natexlab{b}.
\newblock \showarticletitle{{NAIS:} Neural Attentive Item Similarity Model for
  Recommendation}.
\newblock \bibinfo{journal}{\emph{{IEEE} Trans. Knowl. Data Eng.}}
  \bibinfo{volume}{30}, \bibinfo{number}{12} (\bibinfo{year}{2018}),
  \bibinfo{pages}{2354--2366}.
\newblock


\bibitem[\protect\citeauthoryear{He, Liao, Zhang, Nie, Hu, and Chua}{He
  et~al\mbox{.}}{2017b}]%
        {NeuMF}
\bibfield{author}{\bibinfo{person}{Xiangnan He}, \bibinfo{person}{Lizi Liao},
  \bibinfo{person}{Hanwang Zhang}, \bibinfo{person}{Liqiang Nie},
  \bibinfo{person}{Xia Hu}, {and} \bibinfo{person}{Tat{-}Seng Chua}.}
  \bibinfo{year}{2017}\natexlab{b}.
\newblock \showarticletitle{Neural Collaborative Filtering}. In
  \bibinfo{booktitle}{\emph{Proceedings of the 26th International Conference on
  World Wide Web, {WWW} 2017, Perth, Australia, April 3-7, 2017}}.
  \bibinfo{pages}{173--182}.
\newblock


\bibitem[\protect\citeauthoryear{Hidasi, Quadrana, Karatzoglou, and
  Tikk}{Hidasi et~al\mbox{.}}{2016}]%
        {GRU4RecF}
\bibfield{author}{\bibinfo{person}{Bal{\'{a}}zs Hidasi},
  \bibinfo{person}{Massimo Quadrana}, \bibinfo{person}{Alexandros Karatzoglou},
  {and} \bibinfo{person}{Domonkos Tikk}.} \bibinfo{year}{2016}\natexlab{}.
\newblock \showarticletitle{Parallel Recurrent Neural Network Architectures for
  Feature-rich Session-based Recommendations}. In
  \bibinfo{booktitle}{\emph{Proceedings of the 10th {ACM} Conference on
  Recommender Systems, Boston, MA, USA, September 15-19, 2016}}.
  \bibinfo{pages}{241--248}.
\newblock


\bibitem[\protect\citeauthoryear{Huang, Zhao, Dou, Wen, and Chang}{Huang
  et~al\mbox{.}}{2018}]%
        {KSR}
\bibfield{author}{\bibinfo{person}{Jin Huang}, \bibinfo{person}{Wayne~Xin
  Zhao}, \bibinfo{person}{Hongjian Dou}, \bibinfo{person}{Ji{-}Rong Wen}, {and}
  \bibinfo{person}{Edward~Y. Chang}.} \bibinfo{year}{2018}\natexlab{}.
\newblock \showarticletitle{Improving Sequential Recommendation with
  Knowledge-Enhanced Memory Networks}. In \bibinfo{booktitle}{\emph{{SIGIR}}}.
  \bibinfo{publisher}{{ACM}}, \bibinfo{pages}{505--514}.
\newblock


\bibitem[\protect\citeauthoryear{Huang, He, Gao, Deng, Acero, and Heck}{Huang
  et~al\mbox{.}}{2013}]%
        {DSSM}
\bibfield{author}{\bibinfo{person}{Po{-}Sen Huang}, \bibinfo{person}{Xiaodong
  He}, \bibinfo{person}{Jianfeng Gao}, \bibinfo{person}{Li Deng},
  \bibinfo{person}{Alex Acero}, {and} \bibinfo{person}{Larry~P. Heck}.}
  \bibinfo{year}{2013}\natexlab{}.
\newblock \showarticletitle{Learning deep structured semantic models for web
  search using clickthrough data}. In \bibinfo{booktitle}{\emph{22nd {ACM}
  International Conference on Information and Knowledge Management, CIKM'13,
  San Francisco, CA, USA, October 27 - November 1, 2013}}.
  \bibinfo{pages}{2333--2338}.
\newblock


\bibitem[\protect\citeauthoryear{Hug}{Hug}{2020}]%
        {Surprise}
\bibfield{author}{\bibinfo{person}{Nicolas Hug}.}
  \bibinfo{year}{2020}\natexlab{}.
\newblock \showarticletitle{Surprise: A Python library for recommender
  systems}.
\newblock \bibinfo{journal}{\emph{Journal of Open Source Software}}
  \bibinfo{volume}{5}, \bibinfo{number}{52} (\bibinfo{year}{2020}),
  \bibinfo{pages}{2174}.
\newblock


\bibitem[\protect\citeauthoryear{Juan, Zhuang, Chin, and Lin}{Juan
  et~al\mbox{.}}{2016}]%
        {FFM}
\bibfield{author}{\bibinfo{person}{Yu{-}Chin Juan}, \bibinfo{person}{Yong
  Zhuang}, \bibinfo{person}{Wei{-}Sheng Chin}, {and}
  \bibinfo{person}{Chih{-}Jen Lin}.} \bibinfo{year}{2016}\natexlab{}.
\newblock \showarticletitle{Field-aware Factorization Machines for {CTR}
  Prediction}. In \bibinfo{booktitle}{\emph{Proceedings of the 10th {ACM}
  Conference on Recommender Systems, Boston, MA, USA, September 15-19, 2016}}.
  \bibinfo{pages}{43--50}.
\newblock


\bibitem[\protect\citeauthoryear{Kabbur, Ning, and Karypis}{Kabbur
  et~al\mbox{.}}{2013}]%
        {FISM}
\bibfield{author}{\bibinfo{person}{Santosh Kabbur}, \bibinfo{person}{Xia Ning},
  {and} \bibinfo{person}{George Karypis}.} \bibinfo{year}{2013}\natexlab{}.
\newblock \showarticletitle{{FISM:} factored item similarity models for top-N
  recommender systems}. In \bibinfo{booktitle}{\emph{The 19th {ACM} {SIGKDD}
  International Conference on Knowledge Discovery and Data Mining, {KDD} 2013,
  Chicago, IL, USA, August 11-14, 2013}}. \bibinfo{pages}{659--667}.
\newblock


\bibitem[\protect\citeauthoryear{Kang and McAuley}{Kang and McAuley}{2018}]%
        {SASRec}
\bibfield{author}{\bibinfo{person}{Wang{-}Cheng Kang} {and}
  \bibinfo{person}{Julian~J. McAuley}.} \bibinfo{year}{2018}\natexlab{}.
\newblock \showarticletitle{Self-Attentive Sequential Recommendation}. In
  \bibinfo{booktitle}{\emph{{ICDM}}}. \bibinfo{publisher}{{IEEE} Computer
  Society}, \bibinfo{pages}{197--206}.
\newblock


\bibitem[\protect\citeauthoryear{Ke, Meng, Finley, Wang, Chen, Ma, Ye, and
  Liu}{Ke et~al\mbox{.}}{2017}]%
        {LightGBM}
\bibfield{author}{\bibinfo{person}{Guolin Ke}, \bibinfo{person}{Qi Meng},
  \bibinfo{person}{Thomas Finley}, \bibinfo{person}{Taifeng Wang},
  \bibinfo{person}{Wei Chen}, \bibinfo{person}{Weidong Ma},
  \bibinfo{person}{Qiwei Ye}, {and} \bibinfo{person}{Tie{-}Yan Liu}.}
  \bibinfo{year}{2017}\natexlab{}.
\newblock \showarticletitle{LightGBM: {A} Highly Efficient Gradient Boosting
  Decision Tree}. In \bibinfo{booktitle}{\emph{{NIPS}}}.
  \bibinfo{pages}{3146--3154}.
\newblock


\bibitem[\protect\citeauthoryear{Krichene and Rendle}{Krichene and
  Rendle}{2020}]%
        {sampledmetrics}
\bibfield{author}{\bibinfo{person}{Walid Krichene} {and}
  \bibinfo{person}{Steffen Rendle}.} \bibinfo{year}{2020}\natexlab{}.
\newblock \showarticletitle{On Sampled Metrics for Item Recommendation}. In
  \bibinfo{booktitle}{\emph{{KDD} '20: The 26th {ACM} {SIGKDD} Conference on
  Knowledge Discovery and Data Mining, Virtual Event, CA, USA, August 23-27,
  2020}}. \bibinfo{pages}{1748--1757}.
\newblock


\bibitem[\protect\citeauthoryear{Kula and Maciej}{Kula and Maciej}{2017}]%
        {Spotlight}
\bibfield{author}{\bibinfo{person}{Kula} {and} \bibinfo{person}{Maciej}.}
  \bibinfo{year}{2017}\natexlab{}.
\newblock \bibinfo{title}{Spotlight}.
\newblock
  \bibinfo{howpublished}{\url{https://github.com/maciejkula/spotlight}}.
\newblock


\bibitem[\protect\citeauthoryear{Kula}{Kula}{2015}]%
        {LightFM}
\bibfield{author}{\bibinfo{person}{Maciej Kula}.}
  \bibinfo{year}{2015}\natexlab{}.
\newblock \showarticletitle{Metadata Embeddings for User and Item Cold-start
  Recommendations}. In \bibinfo{booktitle}{\emph{Proceedings of the 2nd
  Workshop on New Trends on Content-Based Recommender Systems co-located with
  9th {ACM} Conference on Recommender Systems (RecSys 2015), Vienna, Austria,
  September 16-20, 2015.}} \emph{(\bibinfo{series}{{CEUR} Workshop
  Proceedings}, Vol.~\bibinfo{volume}{1448})}.
  \bibinfo{publisher}{CEUR-WS.org}, \bibinfo{pages}{14--21}.
\newblock


\bibitem[\protect\citeauthoryear{Li, Ren, Chen, Ren, Lian, and Ma}{Li
  et~al\mbox{.}}{2017}]%
        {NARM}
\bibfield{author}{\bibinfo{person}{Jing Li}, \bibinfo{person}{Pengjie Ren},
  \bibinfo{person}{Zhumin Chen}, \bibinfo{person}{Zhaochun Ren},
  \bibinfo{person}{Tao Lian}, {and} \bibinfo{person}{Jun Ma}.}
  \bibinfo{year}{2017}\natexlab{}.
\newblock \showarticletitle{Neural Attentive Session-based Recommendation}. In
  \bibinfo{booktitle}{\emph{Proceedings of the 2017 {ACM} on Conference on
  Information and Knowledge Management, {CIKM} 2017, Singapore, November 06 -
  10, 2017}}. \bibinfo{pages}{1419--1428}.
\newblock


\bibitem[\protect\citeauthoryear{Lian, Zhou, Zhang, Chen, Xie, and Sun}{Lian
  et~al\mbox{.}}{2018}]%
        {xDeepFM}
\bibfield{author}{\bibinfo{person}{Jianxun Lian}, \bibinfo{person}{Xiaohuan
  Zhou}, \bibinfo{person}{Fuzheng Zhang}, \bibinfo{person}{Zhongxia Chen},
  \bibinfo{person}{Xing Xie}, {and} \bibinfo{person}{Guangzhong Sun}.}
  \bibinfo{year}{2018}\natexlab{}.
\newblock \showarticletitle{xDeepFM: Combining Explicit and Implicit Feature
  Interactions for Recommender Systems}. In
  \bibinfo{booktitle}{\emph{Proceedings of the 24th {ACM} {SIGKDD}
  International Conference on Knowledge Discovery {\&} Data Mining, {KDD} 2018,
  London, UK, August 19-23, 2018}}. \bibinfo{pages}{1754--1763}.
\newblock


\bibitem[\protect\citeauthoryear{Liang, Krishnan, Hoffman, and Jebara}{Liang
  et~al\mbox{.}}{2018}]%
        {MultiVAE}
\bibfield{author}{\bibinfo{person}{Dawen Liang}, \bibinfo{person}{Rahul~G.
  Krishnan}, \bibinfo{person}{Matthew~D. Hoffman}, {and} \bibinfo{person}{Tony
  Jebara}.} \bibinfo{year}{2018}\natexlab{}.
\newblock \showarticletitle{Variational Autoencoders for Collaborative
  Filtering}. In \bibinfo{booktitle}{\emph{{WWW}}}. \bibinfo{publisher}{{ACM}},
  \bibinfo{pages}{689--698}.
\newblock


\bibitem[\protect\citeauthoryear{Liu, Zeng, Mokhosi, and Zhang}{Liu
  et~al\mbox{.}}{2018}]%
        {STAMP}
\bibfield{author}{\bibinfo{person}{Qiao Liu}, \bibinfo{person}{Yifu Zeng},
  \bibinfo{person}{Refuoe Mokhosi}, {and} \bibinfo{person}{Haibin Zhang}.}
  \bibinfo{year}{2018}\natexlab{}.
\newblock \showarticletitle{{STAMP:} Short-Term Attention/Memory Priority Model
  for Session-based Recommendation}. In \bibinfo{booktitle}{\emph{Proceedings
  of the 24th {ACM} {SIGKDD} International Conference on Knowledge Discovery
  {\&} Data Mining, {KDD} 2018, London, UK, August 19-23, 2018}}.
  \bibinfo{pages}{1831--1839}.
\newblock


\bibitem[\protect\citeauthoryear{Lobel, Li, Gao, and Carin}{Lobel
  et~al\mbox{.}}{2020}]%
        {RaCT}
\bibfield{author}{\bibinfo{person}{Sam Lobel}, \bibinfo{person}{Chunyuan Li},
  \bibinfo{person}{Jianfeng Gao}, {and} \bibinfo{person}{Lawrence Carin}.}
  \bibinfo{year}{2020}\natexlab{}.
\newblock \showarticletitle{RaCT: Toward Amortized Ranking-Critical Training
  For Collaborative Filtering}. In \bibinfo{booktitle}{\emph{{ICLR}}}.
  \bibinfo{publisher}{OpenReview.net}.
\newblock


\bibitem[\protect\citeauthoryear{Ma, Kang, and Liu}{Ma et~al\mbox{.}}{2019a}]%
        {HGN}
\bibfield{author}{\bibinfo{person}{Chen Ma}, \bibinfo{person}{Peng Kang}, {and}
  \bibinfo{person}{Xue Liu}.} \bibinfo{year}{2019}\natexlab{a}.
\newblock \showarticletitle{Hierarchical Gating Networks for Sequential
  Recommendation}. In \bibinfo{booktitle}{\emph{{KDD}}}.
  \bibinfo{publisher}{{ACM}}, \bibinfo{pages}{825--833}.
\newblock


\bibitem[\protect\citeauthoryear{Ma, Zhou, Cui, Yang, and Zhu}{Ma
  et~al\mbox{.}}{2019b}]%
        {MacridVAE}
\bibfield{author}{\bibinfo{person}{Jianxin Ma}, \bibinfo{person}{Chang Zhou},
  \bibinfo{person}{Peng Cui}, \bibinfo{person}{Hongxia Yang}, {and}
  \bibinfo{person}{Wenwu Zhu}.} \bibinfo{year}{2019}\natexlab{b}.
\newblock \showarticletitle{Learning Disentangled Representations for
  Recommendation}. In \bibinfo{booktitle}{\emph{NeurIPS}}.
  \bibinfo{pages}{5712--5723}.
\newblock


\bibitem[\protect\citeauthoryear{Meng, McCreadie, Macdonald, Ounis, Liu, Wu,
  Wang, Liang, Liang, Zeng, Liang, and Zhang}{Meng et~al\mbox{.}}{2020}]%
        {Beta-recsys}
\bibfield{author}{\bibinfo{person}{Zaiqiao Meng}, \bibinfo{person}{Richard
  McCreadie}, \bibinfo{person}{Craig Macdonald}, \bibinfo{person}{Iadh Ounis},
  \bibinfo{person}{Siwei Liu}, \bibinfo{person}{Yaxiong Wu},
  \bibinfo{person}{Xi Wang}, \bibinfo{person}{Shangsong Liang},
  \bibinfo{person}{Yucheng Liang}, \bibinfo{person}{Guangtao Zeng},
  \bibinfo{person}{Junhua Liang}, {and} \bibinfo{person}{Qiang Zhang}.}
  \bibinfo{year}{2020}\natexlab{}.
\newblock \showarticletitle{BETA-Rec: Build, Evaluate and Tune Automated
  Recommender Systems}. In \bibinfo{booktitle}{\emph{RecSys 2020: Fourteenth
  {ACM} Conference on Recommender Systems, Virtual Event, Brazil, September
  22-26, 2020}}. \bibinfo{pages}{588--590}.
\newblock


\bibitem[\protect\citeauthoryear{Nguyen and Takasu}{Nguyen and Takasu}{2018}]%
        {NPE}
\bibfield{author}{\bibinfo{person}{ThaiBinh Nguyen} {and}
  \bibinfo{person}{Atsuhiro Takasu}.} \bibinfo{year}{2018}\natexlab{}.
\newblock \showarticletitle{{NPE:} Neural Personalized Embedding for
  Collaborative Filtering}. In \bibinfo{booktitle}{\emph{{IJCAI}}}.
  \bibinfo{publisher}{ijcai.org}, \bibinfo{pages}{1583--1589}.
\newblock


\bibitem[\protect\citeauthoryear{Ning and Karypis}{Ning and Karypis}{2011}]%
        {SLIM}
\bibfield{author}{\bibinfo{person}{Xia Ning} {and} \bibinfo{person}{George
  Karypis}.} \bibinfo{year}{2011}\natexlab{}.
\newblock \showarticletitle{{SLIM:} Sparse Linear Methods for Top-N Recommender
  Systems}. In \bibinfo{booktitle}{\emph{{ICDM}}}. \bibinfo{publisher}{{IEEE}
  Computer Society}, \bibinfo{pages}{497--506}.
\newblock


\bibitem[\protect\citeauthoryear{Pan, Xu, Ruiz, Zhao, Pan, Sun, and Lu}{Pan
  et~al\mbox{.}}{2018}]%
        {FwFM}
\bibfield{author}{\bibinfo{person}{Junwei Pan}, \bibinfo{person}{Jian Xu},
  \bibinfo{person}{Alfonso~Lobos Ruiz}, \bibinfo{person}{Wenliang Zhao},
  \bibinfo{person}{Shengjun Pan}, \bibinfo{person}{Yu Sun}, {and}
  \bibinfo{person}{Quan Lu}.} \bibinfo{year}{2018}\natexlab{}.
\newblock \showarticletitle{Field-weighted Factorization Machines for
  Click-Through Rate Prediction in Display Advertising}. In
  \bibinfo{booktitle}{\emph{Proceedings of the 2018 World Wide Web Conference
  on World Wide Web, {WWW} 2018, Lyon, France, April 23-27, 2018}}.
  \bibinfo{pages}{1349--1357}.
\newblock


\bibitem[\protect\citeauthoryear{Paszke, Gross, Massa, Lerer, Bradbury, Chanan,
  Killeen, Lin, Gimelshein, Antiga, et~al\mbox{.}}{Paszke
  et~al\mbox{.}}{2019}]%
        {PyTorch}
\bibfield{author}{\bibinfo{person}{Adam Paszke}, \bibinfo{person}{Sam Gross},
  \bibinfo{person}{Francisco Massa}, \bibinfo{person}{Adam Lerer},
  \bibinfo{person}{James Bradbury}, \bibinfo{person}{Gregory Chanan},
  \bibinfo{person}{Trevor Killeen}, \bibinfo{person}{Zeming Lin},
  \bibinfo{person}{Natalia Gimelshein}, \bibinfo{person}{Luca Antiga},
  {et~al\mbox{.}}} \bibinfo{year}{2019}\natexlab{}.
\newblock \showarticletitle{Pytorch: An imperative style, high-performance deep
  learning library}. In \bibinfo{booktitle}{\emph{Advances in neural
  information processing systems}}. \bibinfo{pages}{8026--8037}.
\newblock


\bibitem[\protect\citeauthoryear{Qu, Cai, Ren, Zhang, Yu, Wen, and Wang}{Qu
  et~al\mbox{.}}{2016}]%
        {PNN}
\bibfield{author}{\bibinfo{person}{Yanru Qu}, \bibinfo{person}{Han Cai},
  \bibinfo{person}{Kan Ren}, \bibinfo{person}{Weinan Zhang},
  \bibinfo{person}{Yong Yu}, \bibinfo{person}{Ying Wen}, {and}
  \bibinfo{person}{Jun Wang}.} \bibinfo{year}{2016}\natexlab{}.
\newblock \showarticletitle{Product-Based Neural Networks for User Response
  Prediction}. In \bibinfo{booktitle}{\emph{{IEEE} 16th International
  Conference on Data Mining, {ICDM} 2016, December 12-15, 2016, Barcelona,
  Spain}}. \bibinfo{pages}{1149--1154}.
\newblock


\bibitem[\protect\citeauthoryear{Ren, Chen, Li, Ren, Ma, and de~Rijke}{Ren
  et~al\mbox{.}}{2019}]%
        {RepeatNet}
\bibfield{author}{\bibinfo{person}{Pengjie Ren}, \bibinfo{person}{Zhumin Chen},
  \bibinfo{person}{Jing Li}, \bibinfo{person}{Zhaochun Ren},
  \bibinfo{person}{Jun Ma}, {and} \bibinfo{person}{Maarten de Rijke}.}
  \bibinfo{year}{2019}\natexlab{}.
\newblock \showarticletitle{RepeatNet: {A} Repeat Aware Neural Recommendation
  Machine for Session-Based Recommendation}. In
  \bibinfo{booktitle}{\emph{{AAAI}}}. \bibinfo{publisher}{{AAAI} Press},
  \bibinfo{pages}{4806--4813}.
\newblock


\bibitem[\protect\citeauthoryear{Rendle}{Rendle}{2010}]%
        {FM}
\bibfield{author}{\bibinfo{person}{Steffen Rendle}.}
  \bibinfo{year}{2010}\natexlab{}.
\newblock \showarticletitle{Factorization Machines}. In
  \bibinfo{booktitle}{\emph{{ICDM} 2010, The 10th {IEEE} International
  Conference on Data Mining, Sydney, Australia, 14-17 December 2010}}.
  \bibinfo{pages}{995--1000}.
\newblock


\bibitem[\protect\citeauthoryear{Rendle}{Rendle}{2012}]%
        {LibFM}
\bibfield{author}{\bibinfo{person}{Steffen Rendle}.}
  \bibinfo{year}{2012}\natexlab{}.
\newblock \showarticletitle{Factorization Machines with {libFM}}.
\newblock \bibinfo{journal}{\emph{ACM Trans. Intell. Syst. Technol.}}
  \bibinfo{volume}{3}, \bibinfo{number}{3}, Article \bibinfo{articleno}{57}
  (\bibinfo{date}{May} \bibinfo{year}{2012}), \bibinfo{numpages}{22}~pages.
\newblock
\showISSN{2157-6904}


\bibitem[\protect\citeauthoryear{Rendle, Freudenthaler, Gantner, and
  Schmidt{-}Thieme}{Rendle et~al\mbox{.}}{2009}]%
        {BPRMF}
\bibfield{author}{\bibinfo{person}{Steffen Rendle}, \bibinfo{person}{Christoph
  Freudenthaler}, \bibinfo{person}{Zeno Gantner}, {and} \bibinfo{person}{Lars
  Schmidt{-}Thieme}.} \bibinfo{year}{2009}\natexlab{}.
\newblock \showarticletitle{{BPR:} Bayesian Personalized Ranking from Implicit
  Feedback}. In \bibinfo{booktitle}{\emph{{UAI} 2009, Proceedings of the
  Twenty-Fifth Conference on Uncertainty in Artificial Intelligence, Montreal,
  QC, Canada, June 18-21, 2009}}. \bibinfo{pages}{452--461}.
\newblock


\bibitem[\protect\citeauthoryear{Rendle, Freudenthaler, and
  Schmidt{-}Thieme}{Rendle et~al\mbox{.}}{2010}]%
        {FPMC}
\bibfield{author}{\bibinfo{person}{Steffen Rendle}, \bibinfo{person}{Christoph
  Freudenthaler}, {and} \bibinfo{person}{Lars Schmidt{-}Thieme}.}
  \bibinfo{year}{2010}\natexlab{}.
\newblock \showarticletitle{Factorizing personalized Markov chains for
  next-basket recommendation}. In \bibinfo{booktitle}{\emph{Proceedings of the
  19th International Conference on World Wide Web, {WWW} 2010, Raleigh, North
  Carolina, USA, April 26-30, 2010}}. \bibinfo{pages}{811--820}.
\newblock


\bibitem[\protect\citeauthoryear{Richardson, Dominowska, and Ragno}{Richardson
  et~al\mbox{.}}{2007}]%
        {LR}
\bibfield{author}{\bibinfo{person}{Matthew Richardson}, \bibinfo{person}{Ewa
  Dominowska}, {and} \bibinfo{person}{Robert Ragno}.}
  \bibinfo{year}{2007}\natexlab{}.
\newblock \showarticletitle{Predicting clicks: estimating the click-through
  rate for new ads}. In \bibinfo{booktitle}{\emph{Proceedings of the 16th
  International Conference on World Wide Web, {WWW} 2007, Banff, Alberta,
  Canada, May 8-12, 2007}}. \bibinfo{pages}{521--530}.
\newblock


\bibitem[\protect\citeauthoryear{Salah, Truong, and Lauw}{Salah
  et~al\mbox{.}}{2020}]%
        {Cornac}
\bibfield{author}{\bibinfo{person}{Aghiles Salah}, \bibinfo{person}{Quoc-Tuan
  Truong}, {and} \bibinfo{person}{Hady~W. Lauw}.}
  \bibinfo{year}{2020}\natexlab{}.
\newblock \showarticletitle{Cornac: A Comparative Framework for Multimodal
  Recommender Systems}.
\newblock \bibinfo{journal}{\emph{Journal of Machine Learning Research}}
  \bibinfo{volume}{21}, \bibinfo{number}{95} (\bibinfo{year}{2020}),
  \bibinfo{pages}{1--5}.
\newblock


\bibitem[\protect\citeauthoryear{Shenbin, Alekseev, Tutubalina, Malykh, and
  Nikolenko}{Shenbin et~al\mbox{.}}{2020}]%
        {RecVAE}
\bibfield{author}{\bibinfo{person}{Ilya Shenbin}, \bibinfo{person}{Anton
  Alekseev}, \bibinfo{person}{Elena Tutubalina}, \bibinfo{person}{Valentin
  Malykh}, {and} \bibinfo{person}{Sergey~I. Nikolenko}.}
  \bibinfo{year}{2020}\natexlab{}.
\newblock \showarticletitle{RecVAE: {A} New Variational Autoencoder for Top-N
  Recommendations with Implicit Feedback}. In
  \bibinfo{booktitle}{\emph{{WSDM}}}. \bibinfo{publisher}{{ACM}},
  \bibinfo{pages}{528--536}.
\newblock


\bibitem[\protect\citeauthoryear{Song, Shi, Xiao, Duan, Xu, Zhang, and
  Tang}{Song et~al\mbox{.}}{2019}]%
        {AutoInt}
\bibfield{author}{\bibinfo{person}{Weiping Song}, \bibinfo{person}{Chence Shi},
  \bibinfo{person}{Zhiping Xiao}, \bibinfo{person}{Zhijian Duan},
  \bibinfo{person}{Yewen Xu}, \bibinfo{person}{Ming Zhang}, {and}
  \bibinfo{person}{Jian Tang}.} \bibinfo{year}{2019}\natexlab{}.
\newblock \showarticletitle{AutoInt: Automatic Feature Interaction Learning via
  Self-Attentive Neural Networks}. In \bibinfo{booktitle}{\emph{Proceedings of
  the 28th {ACM} International Conference on Information and Knowledge
  Management, {CIKM} 2019, Beijing, China, November 3-7, 2019}}.
  \bibinfo{pages}{1161--1170}.
\newblock


\bibitem[\protect\citeauthoryear{Steck}{Steck}{2019}]%
        {EASE}
\bibfield{author}{\bibinfo{person}{Harald Steck}.}
  \bibinfo{year}{2019}\natexlab{}.
\newblock \showarticletitle{Embarrassingly Shallow Autoencoders for Sparse
  Data}. In \bibinfo{booktitle}{\emph{{WWW}}}. \bibinfo{publisher}{{ACM}},
  \bibinfo{pages}{3251--3257}.
\newblock


\bibitem[\protect\citeauthoryear{Sun, Liu, Wu, Pei, Lin, Ou, and Jiang}{Sun
  et~al\mbox{.}}{2019}]%
        {BERT4Rec}
\bibfield{author}{\bibinfo{person}{Fei Sun}, \bibinfo{person}{Jun Liu},
  \bibinfo{person}{Jian Wu}, \bibinfo{person}{Changhua Pei},
  \bibinfo{person}{Xiao Lin}, \bibinfo{person}{Wenwu Ou}, {and}
  \bibinfo{person}{Peng Jiang}.} \bibinfo{year}{2019}\natexlab{}.
\newblock \showarticletitle{BERT4Rec: Sequential Recommendation with
  Bidirectional Encoder Representations from Transformer}. In
  \bibinfo{booktitle}{\emph{Proceedings of the 28th {ACM} International
  Conference on Information and Knowledge Management, {CIKM} 2019, Beijing,
  China, November 3-7, 2019}}. \bibinfo{pages}{1441--1450}.
\newblock


\bibitem[\protect\citeauthoryear{Sun, Yu, Fang, Yang, Qu, Zhang, and Geng}{Sun
  et~al\mbox{.}}{2020}]%
        {daisyRec}
\bibfield{author}{\bibinfo{person}{Zhu Sun}, \bibinfo{person}{Di Yu},
  \bibinfo{person}{Hui Fang}, \bibinfo{person}{Jie Yang},
  \bibinfo{person}{Xinghua Qu}, \bibinfo{person}{Jie Zhang}, {and}
  \bibinfo{person}{Cong Geng}.} \bibinfo{year}{2020}\natexlab{}.
\newblock \showarticletitle{Are We Evaluating Rigorously? Benchmarking
  Recommendation for Reproducible Evaluation and Fair Comparison}. In
  \bibinfo{booktitle}{\emph{Proceedings of the 14th ACM Conference on
  Recommender Systems}}.
\newblock


\bibitem[\protect\citeauthoryear{Tan, Xu, and Liu}{Tan et~al\mbox{.}}{2016}]%
        {ImprovedGRU4Rec}
\bibfield{author}{\bibinfo{person}{Yong~Kiam Tan}, \bibinfo{person}{Xinxing
  Xu}, {and} \bibinfo{person}{Yong Liu}.} \bibinfo{year}{2016}\natexlab{}.
\newblock \showarticletitle{Improved Recurrent Neural Networks for
  Session-based Recommendations}. In \bibinfo{booktitle}{\emph{Proceedings of
  the 1st Workshop on Deep Learning for Recommender Systems, DLRS@RecSys 2016,
  Boston, MA, USA, September 15, 2016}}. \bibinfo{pages}{17--22}.
\newblock


\bibitem[\protect\citeauthoryear{Tang, Qu, Wang, Zhang, Yan, and Mei}{Tang
  et~al\mbox{.}}{2015}]%
        {LINE}
\bibfield{author}{\bibinfo{person}{Jian Tang}, \bibinfo{person}{Meng Qu},
  \bibinfo{person}{Mingzhe Wang}, \bibinfo{person}{Ming Zhang},
  \bibinfo{person}{Jun Yan}, {and} \bibinfo{person}{Qiaozhu Mei}.}
  \bibinfo{year}{2015}\natexlab{}.
\newblock \showarticletitle{{LINE:} Large-scale Information Network Embedding}.
  In \bibinfo{booktitle}{\emph{{WWW}}}. \bibinfo{publisher}{{ACM}},
  \bibinfo{pages}{1067--1077}.
\newblock


\bibitem[\protect\citeauthoryear{Tang and Wang}{Tang and Wang}{2018}]%
        {Caser}
\bibfield{author}{\bibinfo{person}{Jiaxi Tang} {and} \bibinfo{person}{Ke
  Wang}.} \bibinfo{year}{2018}\natexlab{}.
\newblock \showarticletitle{Personalized Top-N Sequential Recommendation via
  Convolutional Sequence Embedding}. In \bibinfo{booktitle}{\emph{Proceedings
  of the Eleventh {ACM} International Conference on Web Search and Data Mining,
  {WSDM} 2018, Marina Del Rey, CA, USA, February 5-9, 2018}}.
  \bibinfo{pages}{565--573}.
\newblock


\bibitem[\protect\citeauthoryear{van~den Berg, Kipf, and Welling}{van~den Berg
  et~al\mbox{.}}{2017}]%
        {GCMC}
\bibfield{author}{\bibinfo{person}{Rianne van~den Berg},
  \bibinfo{person}{Thomas~N. Kipf}, {and} \bibinfo{person}{Max Welling}.}
  \bibinfo{year}{2017}\natexlab{}.
\newblock \showarticletitle{Graph Convolutional Matrix Completion}.
\newblock \bibinfo{journal}{\emph{CoRR}}  \bibinfo{volume}{abs/1706.02263}
  (\bibinfo{year}{2017}).
\newblock
\showeprint[arxiv]{1706.02263}


\bibitem[\protect\citeauthoryear{Wang, Zhang, Ma, Liu, and Ma}{Wang
  et~al\mbox{.}}{2020b}]%
        {ReChorus}
\bibfield{author}{\bibinfo{person}{Chenyang Wang}, \bibinfo{person}{Min Zhang},
  \bibinfo{person}{Weizhi Ma}, \bibinfo{person}{Yiqun Liu}, {and}
  \bibinfo{person}{Shaoping Ma}.} \bibinfo{year}{2020}\natexlab{b}.
\newblock \showarticletitle{Make It a Chorus: Knowledge- and Time-aware Item
  Modeling for Sequential Recommendation}. In
  \bibinfo{booktitle}{\emph{Proceedings of the 43rd International {ACM} {SIGIR}
  conference on research and development in Information Retrieval, {SIGIR}
  2020, Virtual Event, China, July 25-30, 2020}}. \bibinfo{pages}{109--118}.
\newblock


\bibitem[\protect\citeauthoryear{Wang, Zhang, Wang, Zhao, Li, Xie, and
  Guo}{Wang et~al\mbox{.}}{2018}]%
        {RippleNet}
\bibfield{author}{\bibinfo{person}{Hongwei Wang}, \bibinfo{person}{Fuzheng
  Zhang}, \bibinfo{person}{Jialin Wang}, \bibinfo{person}{Miao Zhao},
  \bibinfo{person}{Wenjie Li}, \bibinfo{person}{Xing Xie}, {and}
  \bibinfo{person}{Minyi Guo}.} \bibinfo{year}{2018}\natexlab{}.
\newblock \showarticletitle{RippleNet: Propagating User Preferences on the
  Knowledge Graph for Recommender Systems}. In
  \bibinfo{booktitle}{\emph{Proceedings of the 27th {ACM} International
  Conference on Information and Knowledge Management, {CIKM} 2018, Torino,
  Italy, October 22-26, 2018}}. \bibinfo{pages}{417--426}.
\newblock


\bibitem[\protect\citeauthoryear{Wang, Zhang, Zhang, Leskovec, Zhao, Li, and
  Wang}{Wang et~al\mbox{.}}{2019c}]%
        {KGNN-LS}
\bibfield{author}{\bibinfo{person}{Hongwei Wang}, \bibinfo{person}{Fuzheng
  Zhang}, \bibinfo{person}{Mengdi Zhang}, \bibinfo{person}{Jure Leskovec},
  \bibinfo{person}{Miao Zhao}, \bibinfo{person}{Wenjie Li}, {and}
  \bibinfo{person}{Zhongyuan Wang}.} \bibinfo{year}{2019}\natexlab{c}.
\newblock \showarticletitle{Knowledge-aware Graph Neural Networks with Label
  Smoothness Regularization for Recommender Systems}. In
  \bibinfo{booktitle}{\emph{Proceedings of the 25th {ACM} {SIGKDD}
  International Conference on Knowledge Discovery {\&} Data Mining, {KDD} 2019,
  Anchorage, AK, USA, August 4-8, 2019}}. \bibinfo{pages}{968--977}.
\newblock


\bibitem[\protect\citeauthoryear{Wang, Zhang, Zhao, Li, Xie, and Guo}{Wang
  et~al\mbox{.}}{2019d}]%
        {MKR}
\bibfield{author}{\bibinfo{person}{Hongwei Wang}, \bibinfo{person}{Fuzheng
  Zhang}, \bibinfo{person}{Miao Zhao}, \bibinfo{person}{Wenjie Li},
  \bibinfo{person}{Xing Xie}, {and} \bibinfo{person}{Minyi Guo}.}
  \bibinfo{year}{2019}\natexlab{d}.
\newblock \showarticletitle{Multi-Task Feature Learning for Knowledge Graph
  Enhanced Recommendation}. In \bibinfo{booktitle}{\emph{The World Wide Web
  Conference, {WWW} 2019, San Francisco, CA, USA, May 13-17, 2019}}.
  \bibinfo{pages}{2000--2010}.
\newblock


\bibitem[\protect\citeauthoryear{Wang, Zhao, Xie, Li, and Guo}{Wang
  et~al\mbox{.}}{2019e}]%
        {KGCN}
\bibfield{author}{\bibinfo{person}{Hongwei Wang}, \bibinfo{person}{Miao Zhao},
  \bibinfo{person}{Xing Xie}, \bibinfo{person}{Wenjie Li}, {and}
  \bibinfo{person}{Minyi Guo}.} \bibinfo{year}{2019}\natexlab{e}.
\newblock \showarticletitle{Knowledge Graph Convolutional Networks for
  Recommender Systems}. In \bibinfo{booktitle}{\emph{The World Wide Web
  Conference, {WWW} 2019, San Francisco, CA, USA, May 13-17, 2019}}.
  \bibinfo{pages}{3307--3313}.
\newblock


\bibitem[\protect\citeauthoryear{Wang, Guo, Lan, Xu, Wan, and Cheng}{Wang
  et~al\mbox{.}}{2015}]%
        {HRM}
\bibfield{author}{\bibinfo{person}{Pengfei Wang}, \bibinfo{person}{Jiafeng
  Guo}, \bibinfo{person}{Yanyan Lan}, \bibinfo{person}{Jun Xu},
  \bibinfo{person}{Shengxian Wan}, {and} \bibinfo{person}{Xueqi Cheng}.}
  \bibinfo{year}{2015}\natexlab{}.
\newblock \showarticletitle{Learning Hierarchical Representation Model for
  NextBasket Recommendation}. In \bibinfo{booktitle}{\emph{{SIGIR}}}.
  \bibinfo{publisher}{{ACM}}, \bibinfo{pages}{403--412}.
\newblock


\bibitem[\protect\citeauthoryear{Wang, Fu, Fu, and Wang}{Wang
  et~al\mbox{.}}{2017}]%
        {DCN}
\bibfield{author}{\bibinfo{person}{Ruoxi Wang}, \bibinfo{person}{Bin Fu},
  \bibinfo{person}{Gang Fu}, {and} \bibinfo{person}{Mingliang Wang}.}
  \bibinfo{year}{2017}\natexlab{}.
\newblock \showarticletitle{Deep {\&} Cross Network for Ad Click Predictions}.
  In \bibinfo{booktitle}{\emph{Proceedings of the ADKDD'17, Halifax, NS,
  Canada, August 13 - 17, 2017}}. \bibinfo{pages}{12:1--12:7}.
\newblock


\bibitem[\protect\citeauthoryear{Wang, He, Cao, Liu, and Chua}{Wang
  et~al\mbox{.}}{2019a}]%
        {KGAT}
\bibfield{author}{\bibinfo{person}{Xiang Wang}, \bibinfo{person}{Xiangnan He},
  \bibinfo{person}{Yixin Cao}, \bibinfo{person}{Meng Liu}, {and}
  \bibinfo{person}{Tat{-}Seng Chua}.} \bibinfo{year}{2019}\natexlab{a}.
\newblock \showarticletitle{{KGAT:} Knowledge Graph Attention Network for
  Recommendation}. In \bibinfo{booktitle}{\emph{Proceedings of the 25th {ACM}
  {SIGKDD} International Conference on Knowledge Discovery {\&} Data Mining,
  {KDD} 2019, Anchorage, AK, USA, August 4-8, 2019}}.
  \bibinfo{pages}{950--958}.
\newblock


\bibitem[\protect\citeauthoryear{Wang, He, Wang, Feng, and Chua}{Wang
  et~al\mbox{.}}{2019b}]%
        {NGCF}
\bibfield{author}{\bibinfo{person}{Xiang Wang}, \bibinfo{person}{Xiangnan He},
  \bibinfo{person}{Meng Wang}, \bibinfo{person}{Fuli Feng}, {and}
  \bibinfo{person}{Tat{-}Seng Chua}.} \bibinfo{year}{2019}\natexlab{b}.
\newblock \showarticletitle{Neural Graph Collaborative Filtering}. In
  \bibinfo{booktitle}{\emph{Proceedings of the 42nd International {ACM} {SIGIR}
  Conference on Research and Development in Information Retrieval, {SIGIR}
  2019, Paris, France, July 21-25, 2019}}. \bibinfo{pages}{165--174}.
\newblock


\bibitem[\protect\citeauthoryear{Wang, Jin, Zhang, He, Xu, and Chua}{Wang
  et~al\mbox{.}}{2020a}]%
        {DGCF}
\bibfield{author}{\bibinfo{person}{Xiang Wang}, \bibinfo{person}{Hongye Jin},
  \bibinfo{person}{An Zhang}, \bibinfo{person}{Xiangnan He},
  \bibinfo{person}{Tong Xu}, {and} \bibinfo{person}{Tat{-}Seng Chua}.}
  \bibinfo{year}{2020}\natexlab{a}.
\newblock \showarticletitle{Disentangled Graph Collaborative Filtering}. In
  \bibinfo{booktitle}{\emph{Proceedings of the 43rd International {ACM} {SIGIR}
  conference on research and development in Information Retrieval, {SIGIR}
  2020, Virtual Event, China, July 25-30, 2020}}. \bibinfo{pages}{1001--1010}.
\newblock


\bibitem[\protect\citeauthoryear{Wu, Sun, He, Wang, and Staniforth}{Wu
  et~al\mbox{.}}{2017}]%
        {NeuRec}
\bibfield{author}{\bibinfo{person}{Bin Wu}, \bibinfo{person}{Zhongchuan Sun},
  \bibinfo{person}{Xiangnan He}, \bibinfo{person}{Xiang Wang}, {and}
  \bibinfo{person}{Jonathan Staniforth}.} \bibinfo{year}{2017}\natexlab{}.
\newblock \bibinfo{title}{NeuRec}.
\newblock \bibinfo{howpublished}{\url{https://github.com/wubinzzu/NeuRec}}.
\newblock


\bibitem[\protect\citeauthoryear{Wu, Tang, Zhu, Wang, Xie, and Tan}{Wu
  et~al\mbox{.}}{2019}]%
        {SRGNN}
\bibfield{author}{\bibinfo{person}{Shu Wu}, \bibinfo{person}{Yuyuan Tang},
  \bibinfo{person}{Yanqiao Zhu}, \bibinfo{person}{Liang Wang},
  \bibinfo{person}{Xing Xie}, {and} \bibinfo{person}{Tieniu Tan}.}
  \bibinfo{year}{2019}\natexlab{}.
\newblock \showarticletitle{Session-Based Recommendation with Graph Neural
  Networks}. In \bibinfo{booktitle}{\emph{The Thirty-Third {AAAI} Conference on
  Artificial Intelligence, {AAAI} 2019, The Thirty-First Innovative
  Applications of Artificial Intelligence Conference, {IAAI} 2019, The Ninth
  {AAAI} Symposium on Educational Advances in Artificial Intelligence, {EAAI}
  2019, Honolulu, Hawaii, USA, January 27 - February 1, 2019}}.
  \bibinfo{pages}{346--353}.
\newblock


\bibitem[\protect\citeauthoryear{Wu, DuBois, Zheng, and Ester}{Wu
  et~al\mbox{.}}{2016}]%
        {CDAE}
\bibfield{author}{\bibinfo{person}{Yao Wu}, \bibinfo{person}{Christopher
  DuBois}, \bibinfo{person}{Alice~X. Zheng}, {and} \bibinfo{person}{Martin
  Ester}.} \bibinfo{year}{2016}\natexlab{}.
\newblock \showarticletitle{Collaborative Denoising Auto-Encoders for Top-N
  Recommender Systems}. In \bibinfo{booktitle}{\emph{{WSDM}}}.
  \bibinfo{publisher}{{ACM}}, \bibinfo{pages}{153--162}.
\newblock


\bibitem[\protect\citeauthoryear{Xiao, Ye, He, Zhang, Wu, and Chua}{Xiao
  et~al\mbox{.}}{2017}]%
        {AFM}
\bibfield{author}{\bibinfo{person}{Jun Xiao}, \bibinfo{person}{Hao Ye},
  \bibinfo{person}{Xiangnan He}, \bibinfo{person}{Hanwang Zhang},
  \bibinfo{person}{Fei Wu}, {and} \bibinfo{person}{Tat{-}Seng Chua}.}
  \bibinfo{year}{2017}\natexlab{}.
\newblock \showarticletitle{Attentional Factorization Machines: Learning the
  Weight of Feature Interactions via Attention Networks}. In
  \bibinfo{booktitle}{\emph{Proceedings of the Twenty-Sixth International Joint
  Conference on Artificial Intelligence, {IJCAI} 2017, Melbourne, Australia,
  August 19-25, 2017}}. \bibinfo{pages}{3119--3125}.
\newblock


\bibitem[\protect\citeauthoryear{Xu, Zhao, Liu, Sheng, Xu, Zhuang, Fang, and
  Zhou}{Xu et~al\mbox{.}}{2019}]%
        {GCSAN}
\bibfield{author}{\bibinfo{person}{Chengfeng Xu}, \bibinfo{person}{Pengpeng
  Zhao}, \bibinfo{person}{Yanchi Liu}, \bibinfo{person}{Victor~S. Sheng},
  \bibinfo{person}{Jiajie Xu}, \bibinfo{person}{Fuzhen Zhuang},
  \bibinfo{person}{Junhua Fang}, {and} \bibinfo{person}{Xiaofang Zhou}.}
  \bibinfo{year}{2019}\natexlab{}.
\newblock \showarticletitle{Graph Contextualized Self-Attention Network for
  Session-based Recommendation}. In \bibinfo{booktitle}{\emph{Proceedings of
  the Twenty-Eighth International Joint Conference on Artificial Intelligence,
  {IJCAI} 2019, Macao, China, August 10-16, 2019}}.
  \bibinfo{pages}{3940--3946}.
\newblock


\bibitem[\protect\citeauthoryear{Xue, Dai, Zhang, Huang, and Chen}{Xue
  et~al\mbox{.}}{2017}]%
        {DMF}
\bibfield{author}{\bibinfo{person}{Hong{-}Jian Xue}, \bibinfo{person}{Xinyu
  Dai}, \bibinfo{person}{Jianbing Zhang}, \bibinfo{person}{Shujian Huang},
  {and} \bibinfo{person}{Jiajun Chen}.} \bibinfo{year}{2017}\natexlab{}.
\newblock \showarticletitle{Deep Matrix Factorization Models for Recommender
  Systems}. In \bibinfo{booktitle}{\emph{Proceedings of the Twenty-Sixth
  International Joint Conference on Artificial Intelligence, {IJCAI} 2017,
  Melbourne, Australia, August 19-25, 2017}}. \bibinfo{pages}{3203--3209}.
\newblock


\bibitem[\protect\citeauthoryear{Ying, Zhuang, Zhang, Liu, Xu, Xie, Xiong, and
  Wu}{Ying et~al\mbox{.}}{2018}]%
        {SHAN}
\bibfield{author}{\bibinfo{person}{Haochao Ying}, \bibinfo{person}{Fuzhen
  Zhuang}, \bibinfo{person}{Fuzheng Zhang}, \bibinfo{person}{Yanchi Liu},
  \bibinfo{person}{Guandong Xu}, \bibinfo{person}{Xing Xie},
  \bibinfo{person}{Hui Xiong}, {and} \bibinfo{person}{Jian Wu}.}
  \bibinfo{year}{2018}\natexlab{}.
\newblock \showarticletitle{Sequential Recommender System based on Hierarchical
  Attention Networks}. In \bibinfo{booktitle}{\emph{{IJCAI}}}.
  \bibinfo{publisher}{ijcai.org}, \bibinfo{pages}{3926--3932}.
\newblock


\bibitem[\protect\citeauthoryear{Yuan, Karatzoglou, Arapakis, Jose, and
  He}{Yuan et~al\mbox{.}}{2019}]%
        {NextItNet}
\bibfield{author}{\bibinfo{person}{Fajie Yuan}, \bibinfo{person}{Alexandros
  Karatzoglou}, \bibinfo{person}{Ioannis Arapakis}, \bibinfo{person}{Joemon~M.
  Jose}, {and} \bibinfo{person}{Xiangnan He}.} \bibinfo{year}{2019}\natexlab{}.
\newblock \showarticletitle{A Simple Convolutional Generative Network for Next
  Item Recommendation}. In \bibinfo{booktitle}{\emph{Proceedings of the Twelfth
  {ACM} International Conference on Web Search and Data Mining, {WSDM} 2019,
  Melbourne, VIC, Australia, February 11-15, 2019}}. \bibinfo{pages}{582--590}.
\newblock


\bibitem[\protect\citeauthoryear{Zhang, Yuan, Lian, Xie, and Ma}{Zhang
  et~al\mbox{.}}{2016b}]%
        {CKE}
\bibfield{author}{\bibinfo{person}{Fuzheng Zhang},
  \bibinfo{person}{Nicholas~Jing Yuan}, \bibinfo{person}{Defu Lian},
  \bibinfo{person}{Xing Xie}, {and} \bibinfo{person}{Wei{-}Ying Ma}.}
  \bibinfo{year}{2016}\natexlab{b}.
\newblock \showarticletitle{Collaborative Knowledge Base Embedding for
  Recommender Systems}. In \bibinfo{booktitle}{\emph{Proceedings of the 22nd
  {ACM} {SIGKDD} International Conference on Knowledge Discovery and Data
  Mining, San Francisco, CA, USA, August 13-17, 2016}}.
  \bibinfo{pages}{353--362}.
\newblock


\bibitem[\protect\citeauthoryear{Zhang, Yao, Sun, and Tay}{Zhang
  et~al\mbox{.}}{2019a}]%
        {DLRS-survey}
\bibfield{author}{\bibinfo{person}{Shuai Zhang}, \bibinfo{person}{Lina Yao},
  \bibinfo{person}{Aixin Sun}, {and} \bibinfo{person}{Yi Tay}.}
  \bibinfo{year}{2019}\natexlab{a}.
\newblock \showarticletitle{Deep Learning Based Recommender System: A Survey
  and New Perspectives}.
\newblock \bibinfo{journal}{\emph{ACM Comput. Surv.}} \bibinfo{volume}{52},
  \bibinfo{number}{1}, Article \bibinfo{articleno}{5} (\bibinfo{date}{Feb.}
  \bibinfo{year}{2019}), \bibinfo{numpages}{38}~pages.
\newblock
\showISSN{0360-0300}
\urldef\tempurl%
\url{https://doi.org/10.1145/3285029}
\showDOI{\tempurl}


\bibitem[\protect\citeauthoryear{Zhang, Zhao, Liu, Sheng, Xu, Wang, Liu, and
  Zhou}{Zhang et~al\mbox{.}}{2019b}]%
        {SASRecF}
\bibfield{author}{\bibinfo{person}{Tingting Zhang}, \bibinfo{person}{Pengpeng
  Zhao}, \bibinfo{person}{Yanchi Liu}, \bibinfo{person}{Victor~S. Sheng},
  \bibinfo{person}{Jiajie Xu}, \bibinfo{person}{Deqing Wang},
  \bibinfo{person}{Guanfeng Liu}, {and} \bibinfo{person}{Xiaofang Zhou}.}
  \bibinfo{year}{2019}\natexlab{b}.
\newblock \showarticletitle{Feature-level Deeper Self-Attention Network for
  Sequential Recommendation}. In \bibinfo{booktitle}{\emph{{IJCAI}}}.
  \bibinfo{publisher}{ijcai.org}, \bibinfo{pages}{4320--4326}.
\newblock


\bibitem[\protect\citeauthoryear{Zhang, Zhao, Liu, Sheng, Xu, Wang, Liu, and
  Zhou}{Zhang et~al\mbox{.}}{2019c}]%
        {FDSA}
\bibfield{author}{\bibinfo{person}{Tingting Zhang}, \bibinfo{person}{Pengpeng
  Zhao}, \bibinfo{person}{Yanchi Liu}, \bibinfo{person}{Victor~S. Sheng},
  \bibinfo{person}{Jiajie Xu}, \bibinfo{person}{Deqing Wang},
  \bibinfo{person}{Guanfeng Liu}, {and} \bibinfo{person}{Xiaofang Zhou}.}
  \bibinfo{year}{2019}\natexlab{c}.
\newblock \showarticletitle{Feature-level Deeper Self-Attention Network for
  Sequential Recommendation}. In \bibinfo{booktitle}{\emph{Proceedings of the
  Twenty-Eighth International Joint Conference on Artificial Intelligence,
  {IJCAI} 2019, Macao, China, August 10-16, 2019}}.
  \bibinfo{pages}{4320--4326}.
\newblock


\bibitem[\protect\citeauthoryear{Zhang, Du, and Wang}{Zhang
  et~al\mbox{.}}{2016a}]%
        {FNN}
\bibfield{author}{\bibinfo{person}{Weinan Zhang}, \bibinfo{person}{Tianming
  Du}, {and} \bibinfo{person}{Jun Wang}.} \bibinfo{year}{2016}\natexlab{a}.
\newblock \showarticletitle{Deep Learning over Multi-field Categorical Data - -
  {A} Case Study on User Response Prediction}. In
  \bibinfo{booktitle}{\emph{Advances in Information Retrieval - 38th European
  Conference on {IR} Research, {ECIR} 2016, Padua, Italy, March 20-23, 2016.
  Proceedings}}. \bibinfo{pages}{45--57}.
\newblock


\bibitem[\protect\citeauthoryear{Zhao, Chen, Wang, Gu, and Wen}{Zhao
  et~al\mbox{.}}{2020}]%
        {comparison}
\bibfield{author}{\bibinfo{person}{Wayne~Xin Zhao}, \bibinfo{person}{Junhua
  Chen}, \bibinfo{person}{Pengfei Wang}, \bibinfo{person}{Qi Gu}, {and}
  \bibinfo{person}{Ji{-}Rong Wen}.} \bibinfo{year}{2020}\natexlab{}.
\newblock \showarticletitle{Revisiting Alternative Experimental Settings for
  Evaluating Top-N Item Recommendation Algorithms}. In
  \bibinfo{booktitle}{\emph{{CIKM} '20: The 29th {ACM} International Conference
  on Information and Knowledge Management, Virtual Event, Ireland, October
  19-23, 2020}}. \bibinfo{publisher}{{ACM}}, \bibinfo{pages}{2329--2332}.
\newblock


\bibitem[\protect\citeauthoryear{Zhao, Huang, and Wen}{Zhao
  et~al\mbox{.}}{2016}]%
        {NNRec}
\bibfield{author}{\bibinfo{person}{Wayne~Xin Zhao}, \bibinfo{person}{Jin
  Huang}, {and} \bibinfo{person}{Ji-Rong Wen}.}
  \bibinfo{year}{2016}\natexlab{}.
\newblock \showarticletitle{Learning distributed representations for
  recommender systems with a network embedding approach}. In
  \bibinfo{booktitle}{\emph{Asia information retrieval symposium}}. Springer,
  \bibinfo{pages}{224--236}.
\newblock


\bibitem[\protect\citeauthoryear{Zheng, Lu, Jiang, Zhang, and Yu}{Zheng
  et~al\mbox{.}}{2018}]%
        {SpectralCF}
\bibfield{author}{\bibinfo{person}{Lei Zheng}, \bibinfo{person}{Chun{-}Ta Lu},
  \bibinfo{person}{Fei Jiang}, \bibinfo{person}{Jiawei Zhang}, {and}
  \bibinfo{person}{Philip~S. Yu}.} \bibinfo{year}{2018}\natexlab{}.
\newblock \showarticletitle{Spectral collaborative filtering}. In
  \bibinfo{booktitle}{\emph{Proceedings of the 12th {ACM} Conference on
  Recommender Systems, RecSys 2018, Vancouver, BC, Canada, October 2-7, 2018}}.
  \bibinfo{pages}{311--319}.
\newblock


\bibitem[\protect\citeauthoryear{Zhou, Mou, Fan, Pi, Bian, Zhou, Zhu, and
  Gai}{Zhou et~al\mbox{.}}{2019}]%
        {DIEN}
\bibfield{author}{\bibinfo{person}{Guorui Zhou}, \bibinfo{person}{Na Mou},
  \bibinfo{person}{Ying Fan}, \bibinfo{person}{Qi Pi}, \bibinfo{person}{Weijie
  Bian}, \bibinfo{person}{Chang Zhou}, \bibinfo{person}{Xiaoqiang Zhu}, {and}
  \bibinfo{person}{Kun Gai}.} \bibinfo{year}{2019}\natexlab{}.
\newblock \showarticletitle{Deep Interest Evolution Network for Click-Through
  Rate Prediction}. In \bibinfo{booktitle}{\emph{The Thirty-Third {AAAI}
  Conference on Artificial Intelligence, {AAAI} 2019, The Thirty-First
  Innovative Applications of Artificial Intelligence Conference, {IAAI} 2019,
  The Ninth {AAAI} Symposium on Educational Advances in Artificial
  Intelligence, {EAAI} 2019, Honolulu, Hawaii, USA, January 27 - February 1,
  2019}}. \bibinfo{pages}{5941--5948}.
\newblock


\bibitem[\protect\citeauthoryear{Zhou, Zhu, Song, Fan, Zhu, Ma, Yan, Jin, Li,
  and Gai}{Zhou et~al\mbox{.}}{2018}]%
        {DIN}
\bibfield{author}{\bibinfo{person}{Guorui Zhou}, \bibinfo{person}{Xiaoqiang
  Zhu}, \bibinfo{person}{Chengru Song}, \bibinfo{person}{Ying Fan},
  \bibinfo{person}{Han Zhu}, \bibinfo{person}{Xiao Ma},
  \bibinfo{person}{Yanghui Yan}, \bibinfo{person}{Junqi Jin},
  \bibinfo{person}{Han Li}, {and} \bibinfo{person}{Kun Gai}.}
  \bibinfo{year}{2018}\natexlab{}.
\newblock \showarticletitle{Deep Interest Network for Click-Through Rate
  Prediction}. In \bibinfo{booktitle}{\emph{Proceedings of the 24th {ACM}
  {SIGKDD} International Conference on Knowledge Discovery {\&} Data Mining,
  {KDD} 2018, London, UK, August 19-23, 2018}}. \bibinfo{pages}{1059--1068}.
\newblock


\bibitem[\protect\citeauthoryear{Zhou, Wang, Zhao, Zhu, Wang, Zhang, Wang, and
  Wen}{Zhou et~al\mbox{.}}{2020}]%
        {S3Rec}
\bibfield{author}{\bibinfo{person}{Kun Zhou}, \bibinfo{person}{Hui Wang},
  \bibinfo{person}{Wayne~Xin Zhao}, \bibinfo{person}{Yutao Zhu},
  \bibinfo{person}{Sirui Wang}, \bibinfo{person}{Fuzheng Zhang},
  \bibinfo{person}{Zhongyuan Wang}, {and} \bibinfo{person}{Ji{-}Rong Wen}.}
  \bibinfo{year}{2020}\natexlab{}.
\newblock \showarticletitle{S3-Rec: Self-Supervised Learning for Sequential
  Recommendation with Mutual Information Maximization}. In
  \bibinfo{booktitle}{\emph{{CIKM} '20: The 29th {ACM} International Conference
  on Information and Knowledge Management, Virtual Event, Ireland, October
  19-23, 2020}}. \bibinfo{publisher}{{ACM}}, \bibinfo{pages}{1893--1902}.
\newblock


\end{thebibliography}

\end{document}